\documentclass[12pt]{extarticle}
\pdfoutput=1 
\usepackage{slashed}
\usepackage{xcolor}
\usepackage{latexsym}
\usepackage{amssymb}
\usepackage{amsmath}
\usepackage{graphicx}
\usepackage{subcaption}
\usepackage{arydshln}
\usepackage{adjustbox}
\usepackage{easybmat}
\usepackage{bbm}
\usepackage{cite}
\usepackage{slashed}
\usepackage{bm}
\usepackage{adjustbox}
\usepackage{booktabs}
\usepackage{multirow} 
\usepackage{rotating}
\usepackage{mathtools}
\usepackage{lscape}
\usepackage{booktabs}
\usepackage{hyperref}
\usepackage{bm}
\usepackage{calrsfs}
\usepackage{upgreek}

\newcommand{\nnum}{{\alpha_1\alpha_2\alpha_3}}
\newcommand{\enum}{{\beta_1\beta_2\beta_3}}
\newcommand{\sumnu}{{\Sigma \alpha_i}}
\newcommand{\sumeta}{{\Sigma \beta_i}}
\newcommand{\dd}{\mathrm{d}}

\newcommand{\oi}{{(o)}}
\newcommand{\e}{{(e)}}

\usepackage{listings}
\usepackage{xcolor}

\definecolor{codegreen}{rgb}{0,0.6,0}
\definecolor{codegray}{rgb}{0.5,0.5,0.5}
\definecolor{codepurple}{rgb}{0.58,0,0.82}
\definecolor{backcolour}{rgb}{0.95,0.95,0.92}

\lstdefinestyle{mystyle}{
    backgroundcolor=\color{backcolour},   
    commentstyle=\color{codegreen},
    keywordstyle=\color{magenta},
    numberstyle=\tiny\color{codegray},
    stringstyle=\color{codepurple},
    basicstyle=\ttfamily\footnotesize,
    breakatwhitespace=false,         
    breaklines=true,                 
    captionpos=b,                    
    keepspaces=true,                 
    numbers=left,                    
    numbersep=5pt,                  
    showspaces=false,                
    showstringspaces=false,
    showtabs=false,                  
    tabsize=2
}

\lstdefinelanguage{Mathematica}{
  morekeywords={
    Sum, Table, Integrate, Module, If, Do, Plot, Simplify,
    Expand, Factor, Solve, DSolve, NIntegrate
  },
  sensitive=true,
  morecomment=[s]{(*}{*)},
  morestring=[b]"
}

\lstdefinestyle{mathematica}{
  language=Mathematica,
  basicstyle=\ttfamily\footnotesize,
  keywordstyle=\color{gray},
  commentstyle=\color{green!50!black},
  stringstyle=\color{red!60!black},
  showstringspaces=false,
  columns=fullflexible,
  keepspaces=true,
  frame=single,
  rulecolor=\color{black!30},
  backgroundcolor=\color{gray!5},
  breaklines=true,
  numbers=none
}

\usepackage{algorithm}
\usepackage{algpseudocode}

\usepackage{bbold}
\usepackage[normalem]{ulem}
\newcommand{\sumint}[1]{{\hbox{$\sum$}\!\!\!\!\!\!\!\int\,}_{\!\!\!\!\!\!\raise-0.2ex\hbox{$\scriptstyle{#1}$}}}
\newcommand{\sumintF}[1]{{\hbox{$\sum$}\!\!\!\!\!\!\!\int\,}_{\!\!\!\!\!\!\raise-0.2ex\hbox{$\scriptstyle{\{#1\}}$}}}

\newcommand{\gammaE}[0]{\gamma_\mathrm{E}}

\begin{document}
\thispagestyle{empty}
\begin{flushright}
\end{flushright}
\vspace{0.8cm}

\begin{center}
{\Large\sc SIRENA --- Sum-Integral REductioN Algorithm}
\vspace{0.8cm}

\textbf{
Luis Gil~\footnote{lgil@ugr.es}, Javier López Miras~\footnote{jlmiras@ugr.es} and Adrián Moreno-Sánchez~\footnote{adri@ugr.es}
}\\
\vspace{1.cm}
{\em {Departamento de F\'isica Te\'orica y del Cosmos, Universidad de Granada, Campus de Fuentenueva, E--18071 Granada, Spain}}\\[0.3cm]
\vspace{0.5cm}
\end{center}
\begin{abstract}
We present \texttt{SIRENA}, a \texttt{Python} and \texttt{C++} implementation of the Laporta algorithm for the automatic reduction of multi-loop sum-integrals via integration-by-parts identities. The method builds on established techniques for zero-temperature Feynman integrals and extends them to finite-temperature quantum field theory by consistently accounting for the Matsubara sum structure. We validate the framework by reproducing several known results from the literature up to 3-loop order, and we further provide, for the first time, reductions for selected 3-loop fermionic sum-integrals. In addition to the package, we derive an analytic factorization formula for arbitrary 2-loop fermionic sum-integrals, extending on a previous result for the bosonic case.
\end{abstract}

\setcounter{footnote}{0}

\newpage

\textbf{PROGRAM SUMMARY}

\vspace{1cm}

\begin{small}
\noindent
{\em Program title:} SIRENA \\
{\em Repository link:} \url{https://github.com/lugima/sirena-ibp} \\
{\em Licensing provisions:}  GNU General Public License v3.0. \\
{\em Programming language:} \texttt{Python} and \texttt{C++}\\
{\em Nature of problem:}
Precision computations in quantum field theory at finite temperature involve large sets of loop sum-integrals, which can be reduced to sets of master sum-integrals. Algorithms to perform this reduction up to high loop orders are well-known, and they involve computer algebra techniques which have been fully automated in the case of regular Feynman integrals. Similar public tools that are optimized and adapted to the unique symmetry properties of sum-integrals are lacking. \\
{\em Solution method:}
The program first identifies equivalence classes of sum-integrals via shift symmetry relations and transforms sum-integrals into the representative of the class they belong to. As its main function, it then implements the Laporta algorithm \cite{Laporta:2000dsw} to identify linearly dependent relations between sum-integrals based on integration-by-parts identities at finite temperature \cite{Nishimura:2012ee}, keeping track of bosonic-fermionic signatures of loop momenta. Following techniques analogous to those presented in \cite{Maierhofer:2017gsa}, independent linearly-independent subsystems are parallelized, brought to a triangular form, and back substitution is applied to solve them as a function of the least complex, master sum-integrals.
\end{small}

\newpage

\tableofcontents

\newpage

\section{Introduction}

Owing to its phenomenological relevance in early-universe cosmology, heavy-ion collisions, and astrophysics, quantum field theory (QFT) at finite temperature ($T$) has become a cornerstone of contemporary physics research. The increasing demand for quantitatively reliable predictions has also advanced the field to unprecedented levels of precision in recent years, which has been particularly notable in hot quantum chromodynamics (QCD)~\cite{Laine:2019uua,Laine:2018lgj,Ghiglieri:2021bom,Navarrete:2024ruu,Gorda:2025cwu,Karkkainen:2025nkz} and in the study of cosmological phase transitions (PTs)~\cite{Brauner:2016fla,Andersen:2017ika,Niemi:2018asa,Gorda:2018hvi,Kainulainen:2019kyp,Gould:2019qek,Niemi:2020hto,Gould:2021ccf,Gould:2021dzl,Schicho:2021gca,Lofgren:2021ogg,Niemi:2021qvp,Camargo-Molina:2021zgz,Niemi:2022bjg,Ekstedt:2022bff,Ekstedt:2022ceo,Ekstedt:2022zro,Biondini:2022ggt,Gould:2022ran,Schicho:2022wty,Gould:2023jbz,Kierkla:2023von,Chala:2024xll,Niemi:2024axp,Qin:2024idc,Niemi:2024vzw,Camargo-Molina:2024sde,Gould:2024jjt,Chakrabortty:2024wto,Kierkla:2025qyz,Chala:2025oul,Bernardo:2025vkz,Chala:2025aiz,Chala:2025xlk,Keus:2025ova,Jahedi:2025yjz,Liu:2025ipj,Li:2025kyo,Annala:2025aci,Bhatnagar:2025jhh,Biekotter:2025npc,Chala:2025cya,Liu:2026ask,Bernardo:2026whs}. In both settings, the most successful framework has been shown to be \textit{dimensional reduction} (DR): the construction of three-dimensional (3D) effective field theories (EFTs) \cite{Ginsparg:1980ef, Appelquist:1981vg} in the high-temperature limit. Within the imaginary time formalism, this approach exploits the separation of scales between hard thermal Matsubara excitations of order $\pi T$ and dynamically generated soft scales $g^n T$, with $n \geq 1$ and $g$ denoting the largest coupling in the theory (see \textit{e.g.} \cite{Gould:2023ovu}). DR has enabled consistent weak-coupling expansions while simultaneously allowing for non-perturbative lattice simulations \cite{Niemi:2020hto,Niemi:2022bjg,Gould:2024chm,Niemi:2024axp}, leading for instance to the prediction of the crossover nature of the electroweak PT in the Standard Model \cite{Kajantie:1996mn, Gurtler:1997hr, Csikor:1998eu}. Notably, recent years have also seen significant progress in the automation of the DR pipeline, with dedicated tools such as \texttt{DRalgo} \cite{Ekstedt:2022bff} and the novel package \texttt{Matchotter} for functional matching at finite temperature \cite{Fuentes-Martin:2026bhr}.

At the technical level, the extension of these perturbative computations in DR to high orders requires the systematic reduction and evaluation of large classes of multi-loop sum-integrals, the thermal counterparts of conventional Feynman integrals in the imaginary-time formalism \cite{Matsubara:1955ws}. A substantial and pedagogical body of literature has been devoted to the development of the necessary computational techniques for this purpose, including the extension of integration-by-parts (IBP) methods to finite temperature \cite{Nishimura:2012ee}, algebraic reduction formulae \cite{Davydychev:2023jto}, tensor reduction \cite{Ghisoiu:2012yk,Navarrete:2022adz}, as well as the identification and evaluation of master sum-integrals (MSI) within dimensional regularization \cite{Arnold:1994eb,Arnold:1994ps,Gynther:2007bw,Andersen:2008bz,Moller:2010xw,Ghisoiu:2012yk,Ghisoiu:2012kn,Moller:2012chx,Schroder:2012hm,Ghisoiu:2013zoj,Ghisoiu:2015uza,Davydychev:2023jto,Davydychev:2022dcw,Seppanen:2025owq}. Although these methods were originally developed in the context of hot QCD---where high-order perturbative results were first obtained---they have proven equally useful in the analysis of cosmological PTs to high loop orders \cite{Chala:2025oul,Navarrete:2025yxy,Bernardo:2026whs}, where analogous sum-integral structures arise.

As acknowledged in the aforementioned works, the conceptual framework underlying these techniques closely parallels that of regular Feynman integrals, where they have been automated in a series of public computer algebra packages (\texttt{Air} \cite{Anastasiou:2004vj}, \texttt{FIRE7} \cite{Smirnov:2025prc}, \texttt{Reduze2} \cite{vonManteuffel:2012np}, \texttt{LiteRed} \cite{Lee:2012cn} and \texttt{Kira3} \cite{Lange:2025fba})\footnote{See \cite{Smirnov:2025dfy} and references therein for a thorough review on the development and historical versions of these packages. Notice also recent tools exploring novel approaches such as syzygy and module intersection techniques (\texttt{NeatIBP}~\cite{Wu:2023upw}) or AI-based methods (\texttt{SAILIR}~\cite{Shih:2026jfe}).}. The main diverging point is the evaluation of MSI, where the Matsubara sum prevents the extension of the known techniques at zero temperature, and has so far forced the study of solutions on a case-by-case basis (see \cite{Moller:2012chx} for an exception). Quite surprisingly, no analogous \textit{public} automated tools exist for the IBP reduction process in the thermal case.

In this work, we bridge the gap between formal developments and their practical implementation by presenting a tool for the automatic reduction of thermal sum-integrals: \texttt{SIRENA}. Our approach is based on the well-known Laporta algorithm \cite{Laporta:2000dsw}, and it draws inspiration from the tools available for regular Feynman integrals, adapting them to thermal sum-integrals with modifications that also account for their bosonic-fermionic distinction. We primarily focus on a specific and particularly relevant class of sum-integrals, namely massless vacuum sum-integrals, which arise in the hard-region expansion \cite{Beneke:1997zp, Fuentes-Martin:2024agf} underlying matching in DR. By systematizing and automating these procedures, \texttt{SIRENA} is intended to facilitate high-precision finite temperature computations and to render them accessible to its ever-growing community, at a time when such advances are both timely and necessary. 

In addition to the package, we derive an analytic formula for the factorization of fermionic 2-loop vacuum sum-integrals, building on a previous result for the bosonic case \cite{Davydychev:2023jto}. With this result, we complete the proof of the algebraic factorization of 2-loop sum-integrals started therein, which effectively removes them from perturbative expansions. This result is significant on its own right and further provides a crucial cross-check of the correctness of \texttt{SIRENA}.

This article is organized as follows. In section~\ref{sec:preliminaries} we introduce the notation and summarize the theoretical background for the techniques which are implemented in the code. We dedicate section~\ref{sec:factorization} to the discussion and proof of the algebraic factorization formula for arbitrary 2-loop sum-integrals. Details of the algorithms employed and their implementation are included in section~\ref{sec:workflow}. In section \ref{sec:install} we explain how to install \texttt{SIRENA}, and a tutorial for its usage is provided in section~\ref{sec:usage}. A series of benchmarks for the performance of the code, and brand-new reductions for some 3-loop fermionic sum-integrals are presented in section~\ref{sec:examples}. We finally conclude in section~\ref{sec:conclusions}.

\section{Preliminaries}
\label{sec:preliminaries}

We work in Euclidean $(d+1)$-spacetime dimensions in the imaginary time formalism \cite{Matsubara:1955ws}. In this framework, Euclidean time is compactified to a circle of radius $1/T$ and momenta are split into temporal and spatial components as $P = (P_0, \mathbf{p}) = ([2 n + \sigma] \pi T, \mathbf{p})$, where $n \in \mathbb{Z}$ is the Matsubara mode number and $\sigma = 0 (1)$ denotes the bosonic (fermionic) signature. Given a pair of Euclidean momenta $P$ and $Q$, their Euclidean scalar product is $P \cdot Q = P_0 Q_0 + \mathbf{p} \cdot \mathbf{q}$, where the scalar product between spatial momenta is the one associated to the 3D Euclidean metric, $\delta_{ij} = \mathrm{diag}(+1, +1, +1)$.

Since Euclidean time is compactified to a circle, Feynman integrals are replaced by sum-integrals. In dimensional regularization, with $d=3-2\epsilon$, and in the $\overline{\mathrm{MS}}$ scheme, we denote
\begin{align}
    \sumint{P} &\equiv T \sum_{n=-\infty}^\infty \int_\mathbf{p}
    \,,&
    \int_\mathbf{p} &\equiv \mu^{3-d} \int \frac{{\rm d}^{d}\mathbf{p}}{(2\pi)^{d}}
    \,,
\end{align}
where $\overline{\mu}$ is the $\overline{\mathrm{MS}}$ scale, defined as $\overline{\mu}^2 \equiv 4 \pi e^{-\gammaE} \mu^2$, and $\gammaE$ is the Euler-Mascheroni constant.

\subsection{Vacuum sum-integrals}

The most generic $L$-loop scalar vacuum-type sum-integral (hereinafter, we shall simply refer to them as \textit{sum-integrals}) can be written as
\begin{equation}
    I_{\alpha_1 \dots \alpha_M ; \sigma_1 \dots \sigma_L}^{\beta_1 \dots \beta_M} \equiv \sumint{P_1 \dots P_L} \frac{N_1^{\beta_1} \dots N_M^{\beta_M}}{D_1^{\alpha_1} \dots D_M^{\alpha_M}} \,.
    \label{eq:generic sumint}
\end{equation}
While there exist many possible choices, we choose to use the following \textit{integral family} of massless propagators $D_i$, defined in terms of the $M=L(L+1)/2$ combinations of single and pairs of loop momenta squared
\begin{equation}
    \{ D_1, \dots, D_L, D_{L+1}, \dots, D_M \} = \{ P_1^2, \dots, P_L^2, (P_1-P_2)^2, \dots, (P_{L-1}-P_L)^2 \} \,,
\end{equation}
and an analogous family for combinations $N_i$ of temporal momenta in numerators
\begin{equation}
    \{ N_1, \dots, N_L, N_{L+1}, \dots, N_M \} = \{ P_{0,1}, \dots, P_{0,L}, (P_{0,1}-P_{0,2}), \dots, (P_{0,L-1}-P_{0,L}) \} \,,
\end{equation}
with powers $\alpha_j \in \mathbb{Z}$ and $\beta_j \in \mathbb{N}_0$, and signatures $\sigma_\ell \in \{0, 1\}$ for each loop momentum. Generally, we will drop the $\beta_i$ indices if $\beta_j = 0 ~\forall j$, and we will also omit the $\sigma_k$ indices in fully bosonic integrals ($\sigma_\ell = 0 ~\forall \ell$).

Sum-integrals of the form in eq.~\eqref{eq:generic sumint} are ubiquitous in DR, as they constitute the building blocks of the perturbative expansion of any Green's function computed in a hard-region expansion\footnote{This is also commonly referred to as \textit{soft momentum expansion} in the DR literature.}\cite{Beneke:1997zp}. In this expansion, one assumes that masses $m$ and external momenta $K$ are small compared to loop momenta $P$, so one performs the following series expansion up to a fixed order:
\begin{equation}
    \frac{1}{(P+K)^2 + m^2} = \frac{1}{P^2}\left[1-\frac{K^2+2 P \cdot K + m^2}{P^2} + \dots \right]\,.
\end{equation}
Tensor structures generated in this expansion or, before that, in diagrams involving gauge boson propagators or fermion loops, can be readily reduced to $\mathrm{SO}(3)$ scalars upon simplifying Dirac traces and applying the tensor reduction techniques described in refs. \cite{Ghisoiu:2012yk,Navarrete:2022adz}. Furthermore, leftover scalar products of loop momenta in numerators can always be written as combinations of the $D_i$, \textit{e.g.}
\begin{equation}
    P_1 \cdot P_2 = \frac{1}{2} \left( P_1^2 + P_2^2 - (P_1 - P_2)^2 \right) = \frac{1}{2} \left( D_1^2 + D_2^2 - D_{L+1}^2 \right)\,,
\end{equation}
and analogously, products of 3-momenta can be expressed as
\begin{equation}
    \mathbf{p}_1 \cdot \mathbf{p}_2 = P_1 \cdot P_2 - P_{0,1} P_{0,2} = \frac{1}{2} \left( D_1^2 + D_2^2 - D_{L+1}^2 \right) - N_1 N_2 \,.
\end{equation}

It is clear that the number of different sum-integrals of the form in eq.~\eqref{eq:generic sumint} grows rapidly with increasing loop orders but, fortunately, not all of them are independent. Finding relations between them, and doing so efficiently, is the primary goal of any integral reduction algorithm. These algorithms exploit the many symmetries of these objects to find a minimal (in the sense that they cannot be reduced any further) set of independent integrals, called \textit{masters}, which are conveniently chosen to be the simplest to evaluate. 

As mentioned before, the topic of sum-integral reduction has been covered in the existing literature (see~\cite{Nishimura:2012ee}), so in what follows we shall simply give a brief overview of some fundamental concepts, with a focus on the particular case of sum-integrals.

\subsection{Sector classification}

Once we fix a family of propagators and temporal momenta, any sum-integral of the form in eq.~\eqref{eq:generic sumint} is uniquely identified by three sets of integers:
\begin{equation}
     I_{\alpha_1 \dots \alpha_M ; \sigma_1 \dots \sigma_L}^{\beta_1 \dots \beta_M} \to I((\alpha_1, \dots, \alpha_M), (\beta_1, \dots, \beta_M), (\sigma_1, \dots, \sigma_L)) \equiv I(\vec{\alpha}, \vec{\beta}, \vec{\sigma})\,,
     \label{eq:sint vector}
\end{equation}
that we collect in vectors as short-hand notation. Note that, by restricting to $\beta_j>0$, the components $\beta_{j>L}$ become redundant, as they can be expanded in terms of those with $j\leq L$. Consequently, the general sum-integral may also be written as $I(\vec\alpha, \vec\upbeta, \vec\sigma)$, where $\upbeta = (\upbeta_1, \ldots, \upbeta_L)$. We shall reserve the letters $\beta$ and $\upbeta$ to distinguish between these two representations in section~\ref{sec:canon}.

We define a \textit{sector} as the set of integrals which share the same propagators, \textit{i.e.} for which the subset of non-vanishing $\alpha_j$ is the same. Each of the $2^M$ sectors can be identified with a number $\mathrm{ID}$, whose binary representation is the sequence of 1's and 0's corresponding to present and absent propagators, \textit{i.e.}
\begin{equation}
    \mathrm{ID} = \sum_{j=1}^M (1-\delta_{\alpha_j 0}) 2^{j-1}\,,
\end{equation}
where $\delta_{ab}$ is Kronecker's delta. Note that this differs from the usual definition of sectors, which counts only non-vanishing positive $\alpha_j$ to declare a sector.  A \textit{subsector} of a given sector is another sector whose propagators form a subset of the first.

Following a notation analogous to refs. \cite{vonManteuffel:2012np,Schicho:2020xaf}, we also define the additional identifiers
\begin{align} \label{eq:identifiers ini}
    t &= \sum_{j=1}^M (1-\delta_{\alpha_j 0}) \,, &&\text{number of $D_i$ present}\,, \\
    r &= \sum_{j=1}^M \alpha_j \,, &&\text{sum of $D_i$ powers}\,, \\
    s &= \sum_{j=1}^M \beta_j \,, &&\text{sum of $N_i$ powers}\,, \\
    \alpha_\mathrm{max} &= \mathrm{max} \{\alpha_j\} \,, &&\text{highest $D_i$ power}\,, \\
    \beta_\mathrm{max} &= \mathrm{max} \{\beta_j\} \,, &&\text{highest $N_i$ power}\,,
    \label{eq:identifiers fin}
\end{align}
and we say that all sectors with the same $t$ form a \textit{supersector}. These identifiers provide a measure of the complexity of sum-integrals: larger values correspond to more complex sum-integrals, in the order $\{t, r, s, \alpha_\mathrm{max}, \beta_\mathrm{max}\}$.

Let us note that this choice of ordering prioritizes (\textit{i.e.}, sets as less complex) sum-integrals with fewer denominators. At 3-loop order, employing the usual jargon for sum-integral topologies (see \textit{e.g.} \cite{Ghisoiu:2015uza}), \textit{basketballs} ($t=4$) are favored over \textit{spectacles} ($t=5$) and \textit{Mercedes} ($t=6$). This choice is in a sense convenient, as there exist well-established methods for the evaluation of basketball-type sum-integrals \cite{Arnold:1994ps}, but if the reduction in terms of basketballs involves several divergent prefactors $1/(d-3)^n$, a change of basis might be required to avoid having to evaluate them to higher orders in $\epsilon$. In addition, a reduction in terms of basketballs alone might involve more independent MSI to evaluate, compared to a reduction mixing basketballs and spectacles, so there is no all-encompassing rule for an optimal choice of MSI at this order\footnote{We are indebted to Pablo Navarrete for bringing this to our attention.}. As we indicate in section \ref{sec:usage}, we allow the user to manually choose a set of basis sum-integrals to prioritize as MSI so as to circumvent this potential issue. 

\subsection{Shift symmetries and canonization}
\label{sec:canon}

The invariance of sum-integrals under generic linear momentum shifts of the form
\begin{equation}
    P_i \to \sum_{j=1}^L \mathcal{M}_{ij} P_j
    \label{eq:shifts}
\end{equation}
with $\det(\mathcal{M}) = \pm 1$, and thus unit Jacobian, allows us to find different expressions for the same underlying sum-integral. Taking into account these internal symmetries right before the process of generating a system of IBPs, rather than just appending these extra equations to the latter, drastically reduces its complexity. When performing these shifts, however, the distinction between bosonic and fermionic signatures must be accounted for. Indeed, momentum signatures follow a $\mathbb{Z}_2$ algebra:
\begin{equation}
    P_3 = P_1 \pm P_2 \implies \sigma_3 = \sigma_1 \pm \sigma_2 ~(\mathrm{mod}~ 2)\,,
\end{equation}
so that, in schematic form, 
\begin{align*}
    \text{bosonic} \pm \text{bosonic} &= \text{bosonic}\,, \\
    \text{bosonic} \pm \text{fermionic} &= \text{fermionic}\,, \\
    \text{fermionic} \pm \text{fermionic} &= \text{bosonic}\,.    
\end{align*}
Naturally, when relabeling momentum variables it is also necessary to keep track of their signatures.

Momentum shifts that preserve the integral family structure, \textit{i.e.} those that do not change the bases of $D_i$ or $N_i$, are called \textit{shift symmetries}. These are such that the shift $M$ simply induces a permutation of the indices, that is
\begin{equation}
    I(\vec{\alpha}, \vec{\beta}, \vec{\sigma}) \xrightarrow{\mathcal{M}} I(\vec{\alpha}', \vec{\beta}', \vec{\sigma}')\,; \quad \vec{\alpha}' =  g \vec{\alpha}\,,~ \vec{\beta}' =  g \vec{\beta}\,
\end{equation}
for some $g \in S_M$, and $S_M$ being the group of permutations of $M$ elements (notice that, in general, not all $g \in S_M$ can be realized as momentum shifts). Therefore, shift symmetries induce equivalence relations between sum-integrals in the same family, either within the same sector (\textit{sector symmetries}) or among different sectors (\textit{sector mappings}). This allows us to gather sum-integrals in different equivalence classes.

Through a \textit{canonization} algorithm \cite{vonManteuffel:2012np}, we can find all shift symmetries and exploit them to replace sum-integrals in a given equivalence class by a single representative of said class. Via sector mappings, canonization removes complete sectors in favor of others, whereas sector symmetries allow to relate some sum-integrals in terms of others in the same sector. Therefore, the process of canonization of a particular sum-integral is started with a three-step brute-force algorithm:
\begin{enumerate}
    \item \textbf{Related sectors and possible permutations.} Start with a generic sum-integral belonging to the same sector of the sum-integral at hand, with binary $\mathrm{ID}=b_1 b_2 \ldots b_M$, $b_i \in \{0,1\}$ (the simplest being the \textit{corner sum-integral}, defined as the one satisfying $\alpha_j=b_j, \beta_j = 0, \sigma_\ell = 0$), and apply all possible shifts. The goal of this process is twofold: to determine which sectors are related to this particular one and to store which are the permutations $g\in S_M$ that can be realized via shifts in loop momenta. As a byproduct, we also store the target sector reached by each permutation. 
    
    \item \textbf{Representative sector.} Within the group of sectors in the equivalence class, choose a representative (\textit{e.g.}, the greatest one according to the lexicographic order of their binary IDs) and gather all permutations $G_\mathrm{ID}=\{g\}$ moving from this sector to the representative.
    
    \item \textbf{Representative sum-integral}. Finally, from all the permutations $g$ in $G_\mathrm{ID}$, choose those that lead to the (lexicographically) greater vector $\vec{\alpha}'$.
\end{enumerate}
Notice that the steps 1 and 2, despite being brute-force, only need to be performed once for each sector at the very beginning of the reduction process. Crucially, the process is only required on a few sectors in practice; once a sector is identified with others, all these sectors are in turn identified among themselves, with no need of repeating the process separately for each of them. On the other hand, the last step does need to be carried out for every sum-integral in consideration, but at a much lower computational cost. Visually, the purpose of this procedure is to first order the zeroes in the vector $\vec{\alpha}'$ in a canonical way, and then to reorder the rest of non-zero entries.

At the end of this algorithm $\vec\alpha$ is fixed. However, there could be several $g \in G_\mathrm{ID}$ achieving this, which in turn lead to different $\vec\beta' = g \vec\beta$, while fixing $\vec\alpha$. To choose one of these $\vec\beta'$'s, we follow our canonization picture by transforming each $\vec\beta'$ into a sum of vectors $\vec\upbeta'$ such that $\upbeta'_i=0$ for $i=L+1,\ldots,M$. This is achieved by expanding the zero modes corresponding to sums of loop momenta. Schematically,
\begin{equation}
    P_{0,1}^{\beta'_1} \ldots P_{0,L}^{\beta'_L} (P_{0,1}-P_{0,2})^{\beta'_{L+1}} \ldots (P_{0,L-1}-P_{0,L})^{\beta'_{M}} = \sum_i c_i P_{0,1}^{\upbeta'_{i,1}} \ldots P_{0,L}^{\upbeta'_{i,L}} \,.
    \label{eq:beta expansion}
\end{equation}
Therefore, each permutation $g$ results in a sequence of vectors $\{\vec\upbeta_i\}_g$, which we shall refer to as \textit{expansion of $\upbeta'$'s}. According to the complexity criterion in eqs. \eqref{eq:identifiers ini}-\eqref{eq:identifiers fin}, from all the different expansions of the form \eqref{eq:beta expansion}, given by all different permutations $g \in G_\mathrm{ID}$ keeping $\vec\alpha$ canonical, we choose those such that $\text{max}\,\{\upbeta'_{i,n}\}_g$ is minimum, where $\upbeta'_{i,n}$ are the coordinates of $\vec\upbeta'_i$. If two coincide, we look at the second maximum, and so on. As a last tiebreak, if necessary, we select those permutations which lead to the shortest set $\{\vec\upbeta'_i\}_i$, that is, to fewer terms in the right-hand side of the expansion \eqref{eq:beta expansion}. Then, $\upbeta'$ (or, rather, the expansion of $\upbeta'$'s) is fixed. At this point, the sum-integral is, in general, canonized as a linear combination of several sum-integrals. 

Should the sum-integral present fermionic loop momenta, there could be several permutations $g \in G_\mathrm{ID}$ leading to the same $\vec\alpha'$ and (expansion of) $\vec\upbeta'$, but with different signatures $\vec\sigma'$. To compute the new possible values of $\vec\sigma$ for every $g$, we just keep track of the momentum shift that induces such permutation. By default, we choose as canonical the signature which makes $\vec\sigma'$ minimal (normal lexicographic order). Thus, in a 3-loop case, $(0,0,1)$ is prioritized over $(0,1,0)$ or $(0,1,1)$, for example. An inverse lexicographic order can also be chosen by the user (see section \ref{sec:usage}).

The idea of the canonization algorithm is reminiscent of others already presented in the literature (see \textit{e.g.} \cite{Maierhofer:2017gsa, vonManteuffel:2012np, Smirnov:2025prc}), but its straightforward brute-force approach sets it apart. Moreover, the subtleties regarding the extension to sum-integrals follow the lines of those reviewed in \cite{Schicho:2020xaf}.

\subsection{Integration-by-parts identities}

IBP identities result from the application of the divergence theorem in $d$-dimensions and the fact that Feynman integrals in dimensional regularization are invariant under translations in momentum space, which makes boundary terms vanish \cite{Tkachov:1981wb,Chetyrkin:1981qh}. For sum-integrals, these identities only consider the spatial components of the momenta, and read~\cite{Nishimura:2012ee}
\begin{equation}
    0 = \sumint{P_1 \dots P_L} \frac{\partial}{\partial p_{i, \ell}} \left[ p_{i, \ell'} f(P_1, \dots, P_L, P_{0,1}, \dots, P_{0,L}) \right] \,,
    \label{eq:ibp}
\end{equation}
where $p_{i,\ell}, p_{i,\ell'}$ are loop 3-momenta with $\ell, \ell' \in \{1, \dots, L\}$, and $f$ is a scalar function which we choose to belong to the family of integrands in eq.~\eqref{eq:generic sumint}.  

By running over the different combinations of $\ell, \ell'$, for each integrand $f$ (whose associated integral is called a \textit{seed integral}), this identity generates a set of $L^2$ homogeneous linear equations relating sum-integrals with different $\vec{\alpha}$ and $\vec{\beta}$. Since $T$ is the only scale in massless vacuum sum-integrals, IBP identities provide relations only between sum-integrals with the same mass dimension as the input seed. This is useful to reduce the number of equations to solve for when the mass dimension of the unknown sum-integrals is fixed. Also, crucially, IBP relations do not mix sum-integrals with different signatures.

It is generally observed that by increasing the number of seeds, the number of equations grows faster than the number of unknowns, so there exists a maximum number of seeds one needs to include to reduce any given (sum-)integral to masters. This is the essence of the Laporta algorithm \cite{Laporta:2000dsw}, which has been implemented in several existing codes for regular Feynman integrals \cite{Anastasiou:2004vj,Smirnov:2025prc,vonManteuffel:2012np,Lange:2025fba} (see \textit{e.g.} \cite{Lee:2012cn} for an alternative approach). This approach can be naturally extended to sum-integrals \cite{Nishimura:2012ee}, and it is the one that we follow in the present work. Given the vast literature about it, we refer the interested reader to these references for a thorough description of its implementation.


While IBP identities provide a systematic and general framework for the reduction of multi-loop sum-integrals, they are not always the most economical approach at low loop orders. In particular, extensive evidence shows that all 2-loop sum-integrals admit a factorization into products of 1-loop MSI, and this was proven to be true for the purely bosonic case in \cite{Davydychev:2023jto}, providing an algebraic formula for such factorization. In section~\ref{sec:factorization} we build upon this result and derive a factorization formula for 2-loop fermionic sum-integrals, effectively completing the proof that no 2-loop MSI of the \textit{sunset}-type (namely, non-factorizable) exists. 

\subsection{Even-odd splitting}

There exists an additional relation between fermionic and bosonic sum-integrals that stems from the Matsubara sum \cite{Arnold:1992rz,Coriano:1994re}. Given a bosonic sum-integral $I_{\alpha_1 \dots \alpha_M}^{\beta_1 \dots \beta_M}$ (see eq. \eqref{eq:generic sumint}), if we split each sum in even and odd integers
\begin{equation}
    \sum_{n \in \mathbb{Z}} = \sum_{\substack{n \\ \mathrm{even}}} + \sum_{\substack{n \\ \mathrm{odd}}} 
\end{equation}
and rescale each 3-momenta as $\mathbf{p}_i \to 2 \mathbf{p}_i$, we find the following identity
\begin{equation}
    I_{\alpha_1 \dots \alpha_M}^{\beta_1 \dots \beta_M} = 2^{L d + \sum_{j=1}^M (\beta_j - 2 \alpha_j)} \sum_{\sigma_1=0}^1 \cdots \sum_{\sigma_L=0}^1 I_{\alpha_1 \dots \alpha_M; \sigma_1 \dots \sigma_L}^{\beta_1 \dots \beta_M}\,.
\end{equation}

As a simple example, take the following 2-loop sum-integral
\begin{equation}
    I_{111} = 2^{2d - 3} \left( I_{111} + I_{111;01} + I_{111;10} + I_{111;11} \right) = 2^{2d - 3} \left( I_{111} + 3 I_{111;11} \right)\,, 
\end{equation}
where in the second equality we have used a shift symmetry to rewrite the mixed sum-integrals as fully fermionic. This relation thus determines the fermionic sunset sum-integral as a function of its bosonic counterpart. 

Notably, this identity also allows for any 1-loop fermionic sum-integral to be written as a function of its bosonic counterpart:
\begin{equation}
    I_{\alpha;1}^{\beta} = \left( 2^{2\alpha - \beta - d} - 1 \right) I_{\alpha}^{\beta}\,.
    \label{eq:even-odd 1-loop}
\end{equation}

Note that the simplicity of the examples above depends critically on the fact that the powers in propagators are highly symmetric. In general, these identities introduce large numbers of new mixed sum-integrals at higher-loop order, so their usefulness is restricted to some specific cases. We do not implement these relations in the code, but they can be exploited manually in the IBP-reduced result.

\section{Factorization at 2-loop order}
\label{sec:factorization}

In the context of the reduction strategy outlined in the previous section, 2-loop sum-integrals play a special role. Although they can be treated within the IBP framework implemented in this work, they also allow for a fully algebraic reduction.

It is  observed that, after exhaustive analysis of IBP relations, all 2-loop sum-integrals, independent of their statistics, decompose into products of two 1-loop sum-integrals. This feature, commonly known as \emph{factorization}, was only very recently proven correct for the purely bosonic case~\cite{Davydychev:2023jto}. The authors derived therein an algebraic factorization formula for the factorization of arbitrary 2-loop bosonic sum-integrals, which follows from the corresponding factorization of Feynman integrals with collinear masses at 2-loop order~\cite{Davydychev:2022dcw}. Interestingly, although their fermionic counterparts are also observed to factorize via IBP reduction, no proof or analogous formula were found in said work. In this section we derive an extension of the formula presented therein to the fermionic case, effectively proving that all 2-loop sum-integrals factorize, independent of their signature.

For completeness, we do not only present the final formula, but also its derivation (that it is a minimal modification of the bosonic result). Its implementation as a \texttt{Mathematica} code snippet is provided in appendix \ref{app:snippet}. 

\subsection{Factorization formula}

In the following we shall assume the 2-loop sum-integral family in eq.~\eqref{eq:generic sumint} and use a notation similar to~\cite{Davydychev:2023jto}. 

Assuming positive propagators powers ($\alpha_j \geq 0$), the factorization formula for purely fermionic 2-loop sum-integrals reads
\begin{align}\label{eq: Final Fact Formula}
    I^{\beta_1\beta_2\beta_3}_{\alpha_1\alpha_2\alpha_3;11}=&[1+(-1)^{\Sigma \beta_i}](-1)^{\Sigma\alpha_i}\times\notag\\[0.8 ex]
    &\bigg\{  \sum_{j=1-\alpha_1}^{\alpha_2-1}  c^{({\Sigma \alpha_i})}_{\alpha_1,\alpha_2;j}\sum_{k=0}^{\beta_3}\binom{\beta_3}{k}(-1)^{\beta_2}[1+(-1)^{\ell_1}]F_{\ell_1/2}F_{(2{\Sigma \alpha_i}-{\Sigma \beta_i}-\ell_1)/2}\notag\\[0.8 ex]
    & +\sum_{j=1-\alpha_1}^{\alpha_3-1}  c^{({\Sigma \alpha_i})}_{\alpha_1,\alpha_3;j}\sum_{k=0}^{\beta_2}\binom{\beta_2}{k}(-1)^{\beta_3}[1+(-1)^{\ell_1}]F_{\ell_1/2}I_{(2{\Sigma \alpha_i}-{\Sigma \beta_i}-\ell_1)/2}\notag\\[0.8 ex]
    &+\sum_{j=1-\alpha_2}^{\alpha_3-1}  c^{({\Sigma \alpha_i})}_{\alpha_2,\alpha_3;j}\sum_{k=0}^{\beta_1}\binom{\beta_1}{k}[1+(-1)^{\ell_2}]F_{\ell_2/2}I_{(2{\Sigma \alpha_i}-{\Sigma \beta_i}-\ell_2)/2}\bigg\}\,,
\end{align}
where $\ell_p \equiv \sum_i\alpha_i - \beta_p - j - k$ and we have denoted $I_\alpha\equiv I^0_{\alpha;0}$, $F_\alpha\equiv I^0_{\alpha;1}$, which are related through eq.~\eqref{eq:even-odd 1-loop}. On the other hand, the coefficients $ c^{({\Sigma \alpha_i})}_{\alpha_1,\alpha_2;j}$ are rational functions of the dimension $d$, given by
\begin{align}\label{coefficients}
    &c^{({\Sigma \alpha_i})}_{\alpha_1,\alpha_2;j} \equiv \frac{(-1)^{{\Sigma \alpha_i}-n_j-1}(1-d/2)_{n_j-j-1}}{2(\frac{1}{2})_{n_j-\alpha_2-\alpha_3}(\frac{1}{2})_{n_j-j-\alpha_1-\alpha_3}(\alpha_3-1)!}\times\\[0.8 ex]
    \times&\sum_{k=\max(1+j,1)}^{\min(\alpha_1+j,\alpha_2)}\frac{(d/2-n_j+1)_{k-1}(n_j-k-1)!}{((d+3)/2-{\Sigma \alpha_i})_{n_j-k}(\frac{1}{2})_{\alpha_3-n_j+k}(k-1)!(k-j-1)!(\alpha_2-k)!(\alpha_1-k+j)!}\notag\,,
\end{align}
where we have introduced the Pochhammer symbol $(a)_\alpha\equiv\frac{\Gamma(a+\alpha)}{\Gamma(\alpha)}$ and denoted the integers $n_j \equiv \lceil \frac{{\Sigma\alpha_i}+j}{2}\rceil$. Naturally, the sum-integral $I$ vanishes for odd ${\Sigma\beta_i}$ as one would expect from the definition in eq.~\eqref{eq:generic sumint}, applying the shift $P_{0,p}\to-P_{0,p}$ with $p=1,2$. Furthermore, from the change of variables $P_1 \leftrightarrow P_2$ we obtain the symmetry $ I^{\beta_1\beta_2\beta_3}_{\alpha_1\alpha_2\alpha_2;11}=(-1)^{\beta_3} I^{\beta_2\beta_1\beta_3}_{\alpha_2\alpha_1\alpha_2;11}$ which is present in eq.~\eqref{eq: Final Fact Formula} using the symmetries of the coefficients $ c^{({\Sigma \alpha_i})}_{\alpha_1,\alpha_2;j}$. The factorization formula can be extended for the case with one negative coefficient as follows
\begin{align}
    \label{eq: Final Fact Formula 2}
   &I^{\beta_1\beta_2\beta_3}_{\alpha_1,\alpha_2,\alpha_3\leq0;11}=\frac{1}{2}[1+(-1)^{\Sigma\beta_i}]\sum_{n=0}^{-\alpha_3}\sum_{j=x_2}^{\lfloor(\beta_3+x_2)/2\rfloor}\sum_{k=\max(1,n+{\Sigma\alpha_i}-\alpha_1)}^{\min(\alpha_2,n+{\Sigma\alpha_i}-1)}\sum_{\ell=0}^{\lfloor n/2\rfloor}\times\\[0.8 ex]
    &\times \frac{(-1)^{\beta_2}(-\alpha_3)!2^{2+n-2\ell}\binom{\beta_3}{2j-x}}{\ell!(n-2\ell)!(\alpha_2-k)!(\alpha_1-z)!\Gamma(z)\Gamma(k)}\frac{(d/2)_\ell(1-d/2)_{z-1}(1-d/2)_{k-1}}{(1-z+d/2)_\ell(1-k+d/2)_{\ell}}F_{{\Sigma\alpha_i}-{\Sigma\beta_i}/2-y_2}F_{y_2}\,,\notag\\[5.0 ex]
    \label{eq: Final Fact Formula 3}
    &I^{\beta_1\beta_2\beta_3}_{\alpha_1,\alpha_2\leq0,\alpha_3;11}=\frac{1}{2}[1+(-1)^{\Sigma\beta_i}]\sum_{n=0}^{-\alpha_2}\sum_{j=x_3}^{\lfloor(\beta_2+x_3)/2\rfloor}\sum_{k=\max(1,n+{\Sigma\alpha_i}-\alpha_1)}^{\min(\alpha_3,n+{\Sigma\alpha_i}-1)}\sum_{\ell=0}^{\lfloor n/2\rfloor}\times\\[0.8 ex]
    &\times \frac{(-1)^{\beta_3}(-\alpha_2)!2^{2+n-2\ell}\binom{\beta_2}{2j-x}}{\ell!(n-2\ell)!(\alpha_3
    -k)!(\alpha_1-z)!\Gamma(z)\Gamma(k)}\frac{(d/2)_\ell(1-d/2)_{z-1}(1-d/2)_{k-1}}{(1-z+d/2)_\ell(1-k+d/2)_{\ell}}F_{{\Sigma\alpha_i}-{\Sigma\beta_i}/2-y_3}I_{y_3}\,,\notag
\end{align}
having defined $x_p\equiv \mathrm{mod}(\beta_p+n,2)$, $y_p\equiv k-j-\lfloor (\beta_p+n)/2\rfloor$, and $z\equiv \Sigma_i\alpha_i+n-k$.

Applying the shift $P_1\to P_1-P_2$ yields a relation between the result in eqs.~\eqref{eq: Final Fact Formula}, \eqref{eq: Final Fact Formula 2} and \eqref{eq: Final Fact Formula 3} and 2-loop sum-integrals with mixed bosonic-fermionic momenta. Indeed, for $(\sigma_1\sigma_2) \in \{(10),(01)\}$ we obtain
\begin{align}\label{eq: relations_mixed_momenta}
     I^{\beta_1\beta_2\beta_3}_{\alpha_1\alpha_2\alpha_2;\sigma_1\sigma_2} =  \sigma_1(1-\sigma_2) I^{\beta_1\beta_3\beta_2}_{\alpha_1\alpha_3\alpha_2;11} +  (1-\sigma_1)\sigma_2 (-1)^{\beta_1} I^{\beta_3\beta_2\beta_1}_{\alpha_3\alpha_2\alpha_1;11} \,.
\end{align}

As a final remark, note that eq.~\eqref{eq: Final Fact Formula} closely resembles the bosonic factorization formula of~\cite{Davydychev:2023jto}, the only difference being the signature of the 1-loop sum-integrals involved. Indeed, the bosonic formula is analogous to the one in eq.~\eqref{eq: Final Fact Formula} when we replace all fermionic ($F$) sum-integrals with their bosonic ($I$) counterparts. This is natural from an IBP point of view, since these relations are independent of the signatures. Therefore, one expects the same factorization structure in both cases, the only difference arising from the degeneracy associated with the purely bosonic case. 

All in all, using eqs.~\eqref{eq: Final Fact Formula}, \eqref{eq: Final Fact Formula 2}, \eqref{eq: Final Fact Formula 3} and~\eqref{eq: relations_mixed_momenta}, together with the bosonic result in \cite{Davydychev:2023jto}, any 2-loop sum-integral factorizes into products of 1-loop sum-integrals. 

\subsection{Proof}
\label{app:proof}

The proof of eq.~\eqref{eq: Final Fact Formula} closely follows the steps for the bosonic case (given in detail in \cite{Davydychev:2023jto}). Therefore, in what follows we will focus on the steps where our derivation diverges from the proof in the aforementioned work. Throughout, we will refer to equations appearing there through their numbering (in the journal version), so we  recommend the reader to also follow the original article for a complete version of the proof.

First of all, we observe that doubly-fermionic sum-integrals can be rewritten as a sum of continuum loop integrals over odd integers
\begin{align}\label{eq: F init}
     I^{\enum}_{\nnum;11} &=
\frac{T^2}{(2\pi T)^{2\sumnu - \sumeta - 2d}}
\sum_{n_1,n_2\in \mathbb{Z}}
(n_1+\tfrac{1}{2})^{\eta_1}
(n_2+\tfrac{1}{2})^{\eta_2}
(n_1-n_2)^{\eta_3}
\,B^{\nnum}_{n_1+1/2,n_2+1/2,n_1-n_2}\, \notag
\\[0.8ex]
&=\frac{T^2}{(\pi T)^{2\sumnu-\sumeta-2d}}\,\,\,\sum_{\substack{n_1, n_2 \\ \mathrm{odd}}}n_1^{\eta_1} n_2^{\eta_2} (n_1-n_2)^{\eta_3} \,B^\nnum_{n_1,n_2,n_1-n_2}\,,\\[0.8 ex]
B^{\nnum}_{m_1,m_2,m_3}&\equiv
\int\frac{\dd^d p}{(2\pi)^d}
\int\frac{\dd^d q}{(2\pi)^d}
\frac{1}{
[m_1^2+p^2]^{\alpha_1}
[m_2^2+q^2]^{\alpha_2}
[m_3^2+(p-q)^2]^{\alpha_3}
}\,,
\end{align}
where in the second equality in eq.~\eqref{eq: F init} we have taken out a factor of $2^{- \Sigma\beta_i}$ coming from the sum and another factor of $2^{2\sumnu - 2d}$ stemming from a change of variable in the continuum integral.

We can now decompose the generic integral \eqref{eq: F init} into different summation regions. Compared to the original eq.~(3.4) in the journal version of~\cite{Davydychev:2023jto}, certain terms are absent. This is due to the fact that several regions of the $(n_1,n_2)$-plane are no longer accessible once $n_1,n_2$ are restricted to odd integers. In particular, the integer configurations $(n_1,n_2,n_1-n_2) \in {(0,0,0), (n_1,0,n_1), (0,n_2,-n_2)}$ cannot be realized. Thus, we are  left with
\begin{align}\label{eq: F split}
    I^\enum_{\nnum;11}=\frac{T^2[1+(-1)^\sumnu]}{(\pi T)^{2\sumnu-\sumeta-2d}}\bigg\{&\zeta^\oi(2\sumnu-\sumeta-2d)\times[\delta_{\eta_30}B_{1,1,0}^{\nnum}+2^{\eta_3}B_{1,1,2}^\nnum]\\[0.8ex]
    &+\bar{H}^{\nu_3\nu_2\nu_1}_{\eta_3\eta_2\eta_1}+(-1)^{\eta_3}\bar{H}^{\nu_3\nu_1\nu_2}_{\eta_3\eta_1\eta_2}+(-1)^{\eta_2}H^{\nu_2\nu_1\nu_3}_{\eta_2\eta_1\eta_3}+(-1)^{\eta_2}H^{\nu_1\nu_2\nu_3}_{\eta_1\eta_2\eta_3}\bigg \}\,,
    \notag
\end{align}
where a subscript $(o/e)$ implies that the sum in the zeta function $\zeta(s)$ is performed only over odd/even integers, namely
\begin{equation}
\zeta^\oi(s) \equiv \sum_{n\in\mathbb{N}}\,(2n+1)^{-s}\,, 
\qquad 
\zeta^\e (s) \equiv \sum_{n\in\mathbb{N}}\,(2n)^{-s}\,,
\end{equation}
and we denote
\begin{align}\label{eq: H def}
    H^\nnum_\enum\equiv\sum_{\substack{n_1 > n_2 \\ \mathrm{odd}}}n_1^{\eta_1}n_2^{\eta_2}(n_1+n_2)^{\eta_3}B^\nnum_{n_1,n_2,n_1+n_2}\,,
    \\
    \bar{H}^\nnum_\enum\equiv\sum_{\substack{n_1 > n_2 \\ \mathrm{odd}}}(n_1-n_2)^{\eta_1}n_2^{\eta_2}n_1^{\eta_3}B^\nnum_{n_1-n_2,n_2,n_1}.
    \label{eq: Hbar def}
\end{align}

The continuum integrals $B$ contain what are known as \textit{collinear} masses, \textit{i.e.} they satisfy $m_3=m_1+m_2$, and in this case they can be solved analytically following \cite{Davydychev:2022dcw}. This allows us to rewrite the second line in eq.~\eqref{eq: F split}, in an way similar to the original paper. The only subtlety is that the sums appearing here are over odd $(o)$ and even $(e)$ instead, which yields the following expression
\begin{align}
\label{eq: Sum with all}
    &\frac{\bar{H}^{\nu_3\nu_2\nu_1}_{\eta_3\eta_2\eta_1}+(-1)^{\eta_3}\bar{H}^{\nu_3\nu_1\nu_2}_{\eta_3\eta_1\eta_2}+(-1)^{\eta_2}[H^{\nu_2\nu_1\nu_3}_{\eta_2\eta_1\eta_3}+H^{\nu_1\nu_2\nu_3}_{\eta_1\eta_2\eta_3}]}{(-1)^\sumnu B^{110}_{110}}\notag\\[0.8 ex]
    &=\sum_{j=1-\nu_1}^{\nu_2-1}  c^{(\sumnu)}_{\nu_1,\nu_2;j}\sum_{k=0}^{\eta_3}\binom{\eta_3}{k}((-1)^{\ell_3}+(-1)^{\eta_2})[\zeta^\oi( e_1)\zeta^\oi( d_1)-\zeta^\oi(e_1+d_1)]\notag\\[0.8 ex]
    & +\sum_{j=1-\nu_1}^{\nu_3-1}  c^{(\sumnu)}_{\nu_1,\nu_3;j}\sum_{k=0}^{\eta_2}\binom{\eta_2}{k}\left\{((-1)^{\ell_2}+(-1)^{\eta_3})\zeta^\e(e_1)\zeta^\oi(d_1)-(-1)^{\ell_2}2^{-e_1}\zeta^\oi(e_1+d_1)\right\}\notag\\[0.8 ex]
    &+\sum_{j=1-\nu_2}^{\nu_3-1}  c^{(\sumnu)}_{\nu_2,\nu_3;j}\sum_{k=0}^{\eta_1}\binom{\eta_1}{k}\left\{((-1)^{\ell_2}+1)\zeta^\e(e_2)\zeta^\oi(d_2)-(-1)^{\ell_2}2^{-e_2}\zeta^\oi(e_2+d_2)\right\}\,.
\end{align}
Here $d_p \equiv \ell_p-d$ and $e_p \equiv 2\sumnu-d-\sumeta-\ell_p$. Notice that even summations now appear since some shifts of the kind $n_1 \to n_1\pm n_2$ are needed, thus transforming $n_1$ from odd to even.

Similarly to the discussion around eq.~(5.9) in the original paper, it can be proven that the single zeta functions cancel against the first line of eq.~\eqref{eq: F split} \footnote{There are missing $\zeta(e_1+d_1)$ and $\zeta(e_2+d_2)$ factors in the second and third line of eq.~\eqref{eq: Sum with all} compared to eq.~(5.7) in the original paper. This is  due to a major difference in the so-called \textit{shuffle identity}. The distinction between odd and even summations extends eq.~(A.6) therein to three different cases:
$$
\zeta^\oi(s_1)\zeta^\oi(s_2) = \zeta^{(o,o)}(s_1,s_2) + \zeta^{(o,o)}(s_2,s_1) + \zeta^\oi(s_1+s_2),
\qquad \text{same replacing } \oi\to\e
$$
and
$$
\zeta^\oi(s_1)\zeta^\e(s_2) = \zeta^{(o,e)}(s_1,s_2) + \zeta^{(e,o)}(s_2,s_1)\,,
$$
where $\zeta(s_1,s_2) \equiv \sum_{n_1>n_2>0} n_1^{-s_1} n_2^{-s_2}$ and the superscripts constraint $n_1$ and $n_2$ to only be odd $\oi$ or even $\e$, respectively.
Crucially, the missing term in the last shuffle identity is the one that ultimately would cancel the disappeared $B$ integrals in eq.~\eqref{eq: F init}.}. This leaves us with only products of zeta functions over odd and even numbers. These can be rearranged as follows
\begin{align}\label{app. relationoddvshurtiwitz}
    \zeta^\e(s)=2^{-s}\zeta(s),\qquad \zeta^\oi(s)=2^{-s}\zeta_H(s,1/2)\,, 
\end{align}
where $\zeta_H(s,a) \equiv \sum_{n>0} (n+a)^{-s}$ is the Hurtwitz zeta function (notice that $\zeta_H(s,1/2)$ is nothing but a ``Riemann zeta function with the summation  over positive half-integers''). Finally, the products of Riemann zeta functions can be directly traded for 1-loop bosonic sum-integrals, while the Hurwitz zeta function can be replaced by fermionic ones, as in the original eq.~(6.1). This completes the proof of eq.~\eqref{eq: Final Fact Formula}.

In order to extend the proof to negative powers, we follow the same reasoning as in \cite{Davydychev:2023jto} in eq.~(6.5), since this part of the derivation does not depend on the signature of the sum-integrals. The only particularity is that while the symmetry $I^{\beta_1\beta_2\beta_3}_{\alpha_1\alpha_2\alpha_3;11}=(-1)^{\beta_3}I^{\beta_2\beta_1\beta_3}_{\alpha_2\alpha_1\alpha_3;11}$ is preserved, the symmetry $I^{\beta_1\beta_2\beta_3}_{\alpha_1\alpha_2\alpha_3;11}\neq I^{\beta_3\beta_2\beta_1}_{\alpha_3\alpha_2\alpha_1;11}$ does not hold anymore. For this reason, we have rewritten the result for these two different cases and changed accordingly the signature of the 1-loop sum-integrals involved.

\section{Code workflow}
\label{sec:workflow}

In this section we provide a brief outline of the workflow of \texttt{SIRENA}, which implements many of the techniques already presented in \texttt{Kira} \cite{Maierhofer:2017gsa} with a few exceptions. Users who are mainly interested in a tutorial for the code usage may skip this section and jump directly to section \ref{sec:usage}.

\subsection{Generating initial seeds and IBP relations}

The first step in the reduction process is the generation of a canonization dictionary, which contains all the necessary shift symmetries and sector mappings to express any sum-integral in canonical form, following section \ref{sec:canon}. This computation, which implies testing large numbers of permutations on each sector, can be optionally parallelized in \texttt{SIRENA}, employing several CPU cores via \texttt{Python}'s built-in \texttt{multiprocessing} module, at the user's discretion. The canonization dictionary is only computed once per reduction, and it is thenceforth applied to each sum-integral appearing in the reduction process to canonize it\footnote{This algorithm's performance could be improved by saving the canonization dictionary at each loop order as a file, thereby eliminating the need to re-compute it in each run. However, we observe that this would only reduce computation times significantly beyond the 4-loop order.}.

This is followed by the generation of a set of seeds from which to build a system of IBP relations, a process commonly referred to as \textit{seeding}. Although this step is also only performed once per reduction, the generation of a large set of initial seeds and its subsequent canonization is computationally demanding. Contrary to regular Feynman integrals, however, only sum-integrals of the same mass dimension are related through IBP identities. Thus, given an input set of sum-integrals to reduce, we restrict the seeding to sum-integrals of the same mass dimensions. This vastly reduces the number of seeds needed to reduce the input. Next, canonization is performed independently on each seed, so it can also be parallelized in order to reduce computation time.

For each of the canonized seeds, the corresponding IBP relations are generated as shown in eq.~\eqref{eq:ibp} and the sum-integrals appearing in them are also canonized. IBP equations are stored as \texttt{Python} dictionaries: keys correspond to sum-integrals, their values are their coefficients as $\mathbb{Q}(d)$ polynomials supported by the library \texttt{python-flint}. They represent linear combinations of sum-integrals equaled to zero. All dictionaries are stored in a list, representing a homogeneous, linear system of IBP equations to be reduced.

\subsection{Ordering sum-integrals and equations}

As explained in the original work by Laporta \cite{Laporta:2000dsw}, IBP relations systems are generally very sparse systems of linear equations, but a regular reduction algorithm can spoil this sparsity, thus making the reduction of large systems very inefficient. A smart ordering of the IBP system columns and rows is crucial to avoid the growth in density and minimize the number of substitutions in the reduction process. This technique was originally employed in \texttt{Kira} \cite{Maierhofer:2017gsa}, and it is the same that we follow, so we summarize it here for completeness.

Before defining an order for equations, an order must be defined for the canonized sum-integrals appearing in them. Each sum-integral is ordered from more to least complex according to their $\{t, r, s, \alpha_\mathrm{max}, \beta_\mathrm{max}\}$, lexicographically. This allows to tag sum-integrals with integers from $1$ to $N_\mathrm{sints}$, where $N_\mathrm{sints}$ is the number of different canonized sum-integrals appearing in the system, with smaller numbers corresponding to more complex sum-integrals.

Having an ordering for all sum-integrals, IBP equations are sorted as follows: first by their most complex sum-integral, then by their length, and then by next most complex sum-integrals. This is only a total pre-order, as independent equations with the same sum-integrals but different coefficients are regarded as equally complex. This still allows to sort the IBP system and put most complex equations in the first rows.

\subsection{Reducing the IBP system}

\subsubsection{Decoupling into independent subsystems}

The reduction of linear systems of equations is in general a non-parallelizable algebraic problem. Clearly, in order to solve for a given unknown, not only does one need all equations where said unknown appears, but also the tree of equations containing any other unknown related to the first through any equation. This forces reduction algorithms to consider, in principle, the whole system of equations in order to solve for a given set of unknowns.

Interestingly, however, given the sparsity of large IBP systems, we observe that there do exist completely decoupled subsystems within them which do not share any common sum-integral. In order to maximize the speed of the reduction process, we implement an algorithm that identifies such subsystems and allows to treat each of them separately (see algorithm \ref{alg:decoup_ibps}).

\begin{algorithm}
\small
\caption{Pseudocode: decoupling linear systems of equations}\label{alg:decoup_ibps}
\begin{algorithmic}
\State $L = \{E_1, \dots, E_n\}$ \Comment{Input set of equations}
\State // Initialize set of subsystems
\State $S = \emptyset$
\For{$i = 1$ to $n$}
    \State $V_i = \{ \text{sum-integrals in } E_i \}$
    \State $R_i = \{ i \}$
    \State append $( V_i, R_i )$ to $S$
\EndFor
\State // Iterative merging
\State $\mathrm{merged} = \mathrm{True}$
\While{merged}
    \State $\mathrm{merged} = \mathrm{False}$
    \State $S_\mathrm{new} = \emptyset$
    \While{$S \neq \emptyset$}
        \State pick first $(V,R)$ in $S$
        \State remove $(V, R)$ from $S$
        \State $S_\mathrm{rest} = \emptyset$
        \For{each $(V', R')$ in $S$}
            \If{$V \cap V' \neq \emptyset$}
                \State $V = V \cup V'$
                \State $R = R \cup R'$
                \State $\mathrm{merged} = \mathrm{True}$
            \Else
                \State append $(V', R')$ to $S_\mathrm{rest}$
            \EndIf
        \EndFor
        \State append $(V, R)$ to $S_\mathrm{new}$
        \State $S = S_\mathrm{rest}$
    \EndWhile
    \State $S = S_\mathrm{new}$
\EndWhile
\State // Return separate and sorted subsystems
\State $L_\mathrm{decoup} = \emptyset$
\For{each $(V, R)$ in $S$}
    \State $L_\mathrm{sub} = \emptyset$
    \State $R_\mathrm{sort} = \mathrm{sort}(R)$
    \For{each $r$ in $R_\mathrm{sort}$}
        \State $L_r$ = $r$-th element of $L$ 
        \State append $L_r$ to $L_\mathrm{sub}$
    \EndFor
    \State append $L_\mathrm{sub}$ to $L_\mathrm{decoup}$
\EndFor
\end{algorithmic}
\end{algorithm}

After breaking up the IBP systems into individual subsystems, the user can choose to parallelize the reduction by assigning each subsystem to a different CPU core. In practice, the system is usually decoupled in a few larger subsystems, several smaller ones and many encompassing only a handful of equations.   

\subsubsection{Removing redundant equations}

Each subsystem still typically forms a large, highly redundant system, with many linearly dependent equations. Before reducing each whole subsystem, removing all redundant equations is essential to minimize computation times.

Our algorithm employs a finite field method (evaluating the only symbolic variable $d$ to some pseudo-random integer) with large (random) prime modular arithmetic. This avoids unnecessary manipulations of symbolic coefficients, which are in general rational functions of $d$, and which are not needed in full form to identify linear dependencies (modulo a large prime). This approach, which follows a Gauss elimination algorithm, is similar to the one implemented in \texttt{Kira} \cite{Maierhofer:2017gsa} in \texttt{C++}, but we carry out our own implementation in \texttt{Python}.

\subsubsection{Gauss type forward elimination and back substitution}
\label{sec:gauss}

The \textit{Gauss type forward elimination algorithm} presented in \texttt{Kira} \cite{Maierhofer:2017gsa}, and that we also employ, is different from standard Gaussian elimination. The subsystem is first sorted and equations are grouped by their most complex sum-integral. Equations in each of these groups are structurally similar, so Gauss elimination does not generate many new terms (it does not increase the density). The process of elimination is engineered to use the most complex equation in order to eliminate the most complex sum-integral from the rest of equations within the group. After elimination is carried out in each separate group, the system is sorted and re-grouped, and the process is carried out again. This way, the system is eventually brought into upper-right triangular form while preserving its sparsity to a great extent. 

Once the system is in triangular form, regular \textit{back substitution} is applied to express the most complex sum-integrals in terms of the simplest, which effectively yields expressions for all sum-integrals in terms of MSI.

We implement these algorithms in \texttt{C++}, which greatly reduces the typical computation time with respect to an initial implementation in \texttt{Python}. The translation between \texttt{C++} and \texttt{Python} types is dealt with using the library \texttt{pybind11} \cite{pybind11}, and the relevant \texttt{C++} functions are imported as a \texttt{Python} module.

\subsubsection{Finding minimal set of masters}
\label{sec:find masters}

One difficulty associated to the Laporta algorithm is that, given a set of sum-integrals to reduce, the minimal set of seeds needed to yield a reduction to the simplest MSI cannot be known \textit{a priori}. Instead, one tries with increasingly large sets of seeds until the reduction is satisfactory. If the input sum-integrals contain negative powers in propagators, this usually implies adding negative powers in the seeds, which can drastically enlarge the set of seeds and make the reduction slow.

To avoid having to manually re-run the reduction with different initial seeds, and with seeds with many negative powers, \texttt{SIRENA} can automatically identify if sum-integrals can be further reduced by adding small sets of neighboring sum-integrals to the original seeds. A \textit{neighbor} of a sum-integral is another sum-integral with the same sum of propagator powers ($r$), but with one propagator raised to one greater power and another raised to one more negative power. Sum-integrals are connected to their neighbors through IBP identities directly, so in principle this approach makes it likely to reduce the set of MSI in a few iterations without enlarging the initial set of seeds too much. The algorithm proceeds as follows:
\begin{enumerate}
    \item[0.] Take the number of MSI to be zero and seeds to be the ones in the initial set.
    \item Generate the ensuing IBP system and reduce it  at finite field value for $d$. By using large prime modular arithmetics, sum-integrals which are not reduced to simpler cases (MSI) are identified reliably.
    \item If the MSI set is the same as in the previous iteration, go to (4). Else, go to (3).
    \item Add all neighbors of the identified MSI to the set of seeds. Go back to (1).
    \item Reduce the IBP system symbolically with rational functions of $d$. End.
\end{enumerate}

Since the symbolic coefficients of the reduction are not required until the complete set of necessary seeds is determined, this re-run process can be performed numerically using modular finite field arithmetic. Once the seeding is optimized, the reduction is performed one last time symbolically. Given that finite field methods provide a significant boost to computational performance, this is the go-to approach in \texttt{SIRENA}.

This procedure does not ensure that the final set of MSI is the simplest according to the chosen complexity criterion, but it aids in getting closer to it without any intervention from the user. 

\section{Installation}
\label{sec:install}

\texttt{SIRENA} is free software under the terms of the GNU General Public License v3.0 and its source code is publicly available in the GitHub repository

\begin{center}
    \url{https://github.com/lugima/sirena-ibp}
\end{center}

In order to run \texttt{SIRENA}, a version of \texttt{Python} \texttt{3.10+} is required. At the time of publication, the code has been tested up to version \texttt{3.14}.

\texttt{SIRENA} is uploaded to the official PyPI repository, so it can be easily installed via \texttt{pip} executing the following command in a terminal:
\begin{lstlisting}[language=sh]
$ pip install sirena-ibp
\end{lstlisting}

Since the package contains \texttt{C++} source files, compilation is required. These are based on the \texttt{FLINT} library \cite{flint} for efficient polynomial algebra manipulation. Nevertheless, the installation process is designed to be automatic and user-transparent. Leveraging \texttt{GitHub Actions}, the project is pre-compiled into platform-specific distributions for \texttt{GNU/Linux}\footnote{We are currently working on extending this compatibility to \texttt{macOS} systems.}. Consequently, a standard \texttt{pip install} from the \texttt{PyPI} repository provides a ready-to-use version of the package that includes all necessary components. Crucially, the end-user is not required to manually install the \texttt{FLINT} library, a \texttt{C++} compiler, or any additional build tools, as these dependencies are already bundled within the distributed package.

In the event that the \texttt{pip install} command fails, or if a manual installation is preferred, the user must have a \texttt{C++} compiler and the \texttt{FLINT} library (v3.0+) pre-installed. If they are not found in the standard system locations, their path must be specified by modifying the search directories in the \texttt{CMakeLists.txt} file located in \texttt{src/sirena/cpp/}. Once configured, manual compilation of the source files is unnecessary; running a local \texttt{pip} installation, via:
\begin{lstlisting}[language=sh]
$ pip install .
\end{lstlisting}
from the project's root directory will handle the build process automatically.

After the installation, as a standard \texttt{Python} package, all \texttt{SIRENA} routines can be imported into any \texttt{Python} environment using:
\begin{lstlisting}[language=Python]
import sirena
\end{lstlisting}
Furthermore, the installation automatically configures a \textit{Command Line Interface} (CLI), making the \texttt{sirena} command available globally directly from the system terminal and bypassing any \texttt{Python} instruction. By executing:
\begin{lstlisting}
$ sirena --help
\end{lstlisting}
a description of the package and the available options are displayed. For more details about the usage of the package we refer the reader to section~\ref{sec:usage}.

Finally, to ensure a clean installation and prevent conflicts with existing system libraries, it is highly recommended to install the package within a dedicated Python virtual environment (\textit{e.g.}, using \texttt{venv} or \texttt{conda}).

\section{Using \texttt{SIRENA}}
\label{sec:usage}

\subsection{Input and options}

In \texttt{SIRENA} sum-integrals are identified with tuples containing three tuples of integers, as anticipated in eq.~\eqref{eq:sint vector}. All operations that the code performs on sum-integrals preserve the integral family, so they can always be described in terms of these tuples.

The user might be interested in finding the reduction of a set of sum-integrals, or a linear combination of them with symbolic coefficients, as they usually appear in perturbative computations. Both cases can be provided as input to \texttt{SIRENA} via an input \texttt{.txt} file. A sample 3-loop input looks like:
\begin{lstlisting}
sints_in = (((a11,a12,a13,a14,a15,a16), (b11,b12,b13,b14,b15,b16), (s11,s12,s13)),((a21,a22,a23,a24,a25,a26), (b21,b22,b23,b24,b25,b26), (s21,s22,s23)), ...)

coeffs_in = (c1, c2, ...)
\end{lstlisting}
which is interpreted as a sum of sum-integrals multiplied by the corresponding coefficients\footnote{Actually, \texttt{bn4,bn5,bn6} (or, generically, the last $M-L$ numerator indices) can be omitted, as they correspond to mixed numerators which are usually expanded in perturbative computations. The code interprets the missing numerator indices as zero.}. \texttt{coeffs\_in} can contain integers, fractions or symbols representing parameters (couplings, dimension $d$, \dots), which are automatically parsed as \texttt{sympy} \cite{sympy} expressions. The coefficients list can be omitted, so only the reduction for each individual sum-integral is obtained. Furthermore, another optional list
\begin{lstlisting}[language=sh]
priority = (((A11,A12,A13,A14,A15,A16), (B11,B12,B13,B14,B15,B16), (S11,S12,S13)), ((A21,A22,A23,A24,A25,A26), (B21,B22,B23,B24,B25,B26), (S21,S22,S23)), ...)
\end{lstlisting}
with any number of sum-integrals can be provided in the same file, which sets the sum-integrals to prioritize as MSI for the reduction. This is useful to express results in different MSI bases beyond 2-loop order, since not all MSI factorize to 1-loop sum-integrals. It is important that the names \texttt{sints\_in}, \texttt{coeffs\_in} and \texttt{priority} are written in this way, and we remind the user that while the first one is compulsory, the other two are optional.

From the input, the code extracts some characteristics of the given sum-integrals. First, it determines whether there are fermionic signatures in the input set. This is relevant because in the case of a fully bosonic input, seed generation can be restricted to bosonic integrals, which greatly reduces the size of the IBP system to solve. Next, it computes the different mass dimensions $[I]$ of the input as
\begin{equation}
    [I] = 4 L + \sum_{j=1}^M (\beta_j - 2 \alpha_j)\,,
\end{equation}
so the initial set of seeds is reduced to only those with the same mass dimension(s). 

Another file, \texttt{params.txt}, can be optionally provided with some useful parameters for the reduction. This file should look like\footnote{The values in the example are the default used in the code if the parameters file is not provided.}:
\begin{lstlisting}
max_r = 6
max_s = 6
alpha_ini = 0
sig_order = "normal"

rerun = True
n_cpus = ["auto", 1]
to_wolfram = True
\end{lstlisting}
The first three parameters fix the maximum $r$ and $s$ (see eqs.~\eqref{eq:identifiers ini} to \eqref{eq:identifiers fin}) and the most negative propagator power to include in the initial seed generation, respectively\footnote{Naturally, \texttt{max\_r} and \texttt{max\_s} must be set to values equal or larger than the corresponding values in the list of input sum-integrals.}. \texttt{sig\_order} indicates which lexicographic ordering should be used for signatures in the canonization process (\texttt{"normal"} or \texttt{"inverse"}; see section \ref{sec:canon}). 

On the other hand, the parameter \texttt{rerun} can be set to \texttt{True} if the reduction should be carried out iteratively until the number of MSI converges via the addition of neighbors (see section \ref{sec:find masters}), or to \texttt{False} if it should be carried out only once. The number of CPU cores to use for seed canonization (first element) and for the solution of the IBP system (second element) is provided in \texttt{n\_cpus}. If set to \texttt{"auto"}, all available cores are used. Finally, \texttt{to\_wolfram} indicates whether the output should be written in a format compatible with Wolfram Language, for an easier port to \texttt{Mathematica}. 

\subsection{\texttt{SIRENA} from the Command Line Interface}

Once installed, to use \texttt{SIRENA} directly from the CLI the user can simply open a terminal and write the following command\footnote{If using a virtual environment, it must be activated before running \texttt{SIRENA} from the terminal.}:
\begin{lstlisting}[language=sh]
$ sirena [-h] [-p PARAMS] [-v] [--demo] INPUT OUTPUT
\end{lstlisting}
The first four are optional arguments: \texttt{-h} shows a help message, \texttt{-p PARAMS} is used to provide the path of the parameters file, \texttt{-v} enables printing the reduction progress on screen and \texttt{--demo} runs a test reduction to check the correct installation. For a more detailed on-screen output, \texttt{-vv} can be used instead of \texttt{-v}\footnote{If more detailed output is activated and the number of cores to use for the solution of the system is larger than one, the information about the solving process for each subsystem is not displayed, in order not to scramble the information coming from different parallelized subsystems.}. On the other hand, \texttt{INPUT} and \texttt{OUTPUT} are paths of the input and output \texttt{.txt} files, respectively, and they must always be provided (unless running \texttt{--demo}). Note that the user can specify file paths as either absolute or relative to the current working directory.

In the output file, \texttt{SIRENA} writes: (i) the reduction of each input sum-integral (\texttt{sints\_out}), (ii) the list of MSI that appear in that reduction (\texttt{masters}) and, if \texttt{coeffs\_in} was provided, (iii) the combined result after the reduction (\texttt{combined}). Sum-integrals are written in a more legible format, similar to the left-hand side of eq.~\eqref{eq:sint vector}. Also, if \texttt{to\_wolfram} is set to \texttt{True} in the parameters file, the output is written in Wolfram Language.

\subsection{\texttt{SIRENA} from a \texttt{Python} interface}

While we encourage the user to use the CLI routine for its simplicity, the instructions can analogously be given within a \texttt{Python} environment. For completeness, we explain the main features of \texttt{sirena} as a \texttt{Python} package.

The main function \texttt{sirena}:
\begin{lstlisting}[language=Python]
from sirena import sirena

sirena(sints, 
        max_r=6, 
        max_s=6, 
        alpha_ini=0, 
        sig_order="normal",
        rerun=True,
        n_cpus=["auto",1], 
        basis_sints=[])
\end{lstlisting}
computes the reduction of the list sum-integrals in the variable \texttt{sints}. The optional arguments \texttt{max\_r}, \texttt{max\_s}, \texttt{alpha\_ini}, \texttt{sig\_order}, \texttt{rerun}, \texttt{n\_cpus} are the options as specified by the parameter file, whereas the \texttt{basis\_sints} argument corresponds to the list of sum-integrals to prioritize, namely, the variable \texttt{priority} in the input file. The output of this routine is a dictionary relating each sum-integral in \texttt{sints} with its expression in terms of the MSI. In turn, the expression itself is saved as a dictionary compassing the MSI with their coefficients ($\mathbb{Q}(d)$ polynomials in \texttt{sympy}).

The solution can be exported to a text file with the \texttt{sints\_to\_text} routine:
\begin{lstlisting}[language=Python]
from sirena import sints_to_txt

sints_to_txt(sints, 
        sols,
        file,
        coeffs_in=None,
        to_wolfram=False)
\end{lstlisting}
It requires the list of sum-integrals to solve for (\texttt{sints}), their solution (\texttt{sols}), as directly returned from the \texttt{sirena} function, and the file name (\texttt{file}) into which the results will be written. If a list of coefficients (again, as \texttt{sympy} expressions) is given, then the linear combination of the sum-integrals with this choice of coefficients will also be written in the output file. Finally, if the argument \texttt{to\_wolfram} is set to \texttt{True}, the output will be formatted accordingly to the Wolfram Language.

The input sum-integrals, coefficients and sum-integrals to prioritize as basis can be directly read from a file, in this specific order, via the function \texttt{sints\_from\_txt(file)}. Likewise, the rest of parameters can be obtained from the parameter file with \linebreak \texttt{params\_from\_txt(file)}. While the first routine returns three arguments, the latter outputs a dictionary with the name of the options as keys.

In this way, a simplified standard \texttt{SIRENA} pipeline could look like the following:
\begin{lstlisting}[language=Python]
from sirena import sirena, sints_to_txt, sints_from_txt, params_from_txt

input_file = "input.txt"
output_file = "output.txt"
params_file = "params.txt"

sints_in, coeffs_in, priority = sints_from_txt(input_file)
params = params_from_txt(params_file)

sols = sirena(sints_in, 
    max_r=params["max_r"], 
    max_s=params["max_s"],
    alpha_ini=params["alpha_ini"],
    sig_order=params["sig_order"],
    rerun=params["rerun"],
    n_cpus=params["n_cpus"], 
    basis_sints=priority)

sints_to_txt(sints_in, 
    sols,
    output_file,
    coeffs_in=coeffs_in, 
    to_wolfram=params["to_wolfram"])
\end{lstlisting}

\section{Benchmarks and examples}
\label{sec:examples}

\begin{figure}[h]
    \centering
    \includegraphics[width=0.6\linewidth]{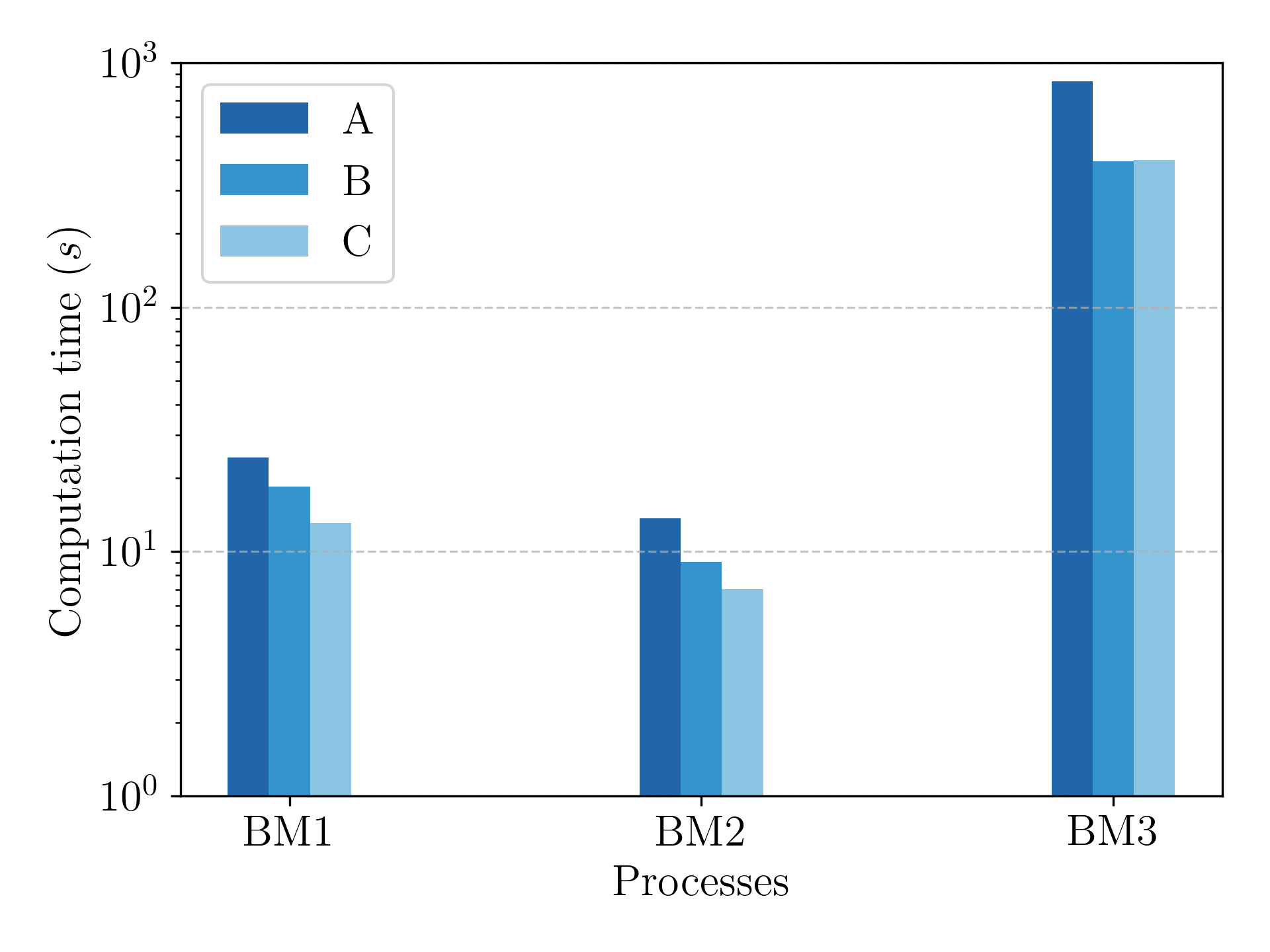}
    \caption{Performance of the different systems (A, B, C) in the three benchmarks. The comparison is performed parallelizing the seed generation and reduction in 20 CPU cores in all systems.}
    \label{fig:barplot}
\end{figure}

To assess the performance of \texttt{SIRENA}, we study a series of reduction examples\footnote{The corresponding input files are included in the program files, in the \texttt{examples/} directory, so we encourage any potential user to use them as reference for their own reductions.} in three different systems: A, B, C. Their hardware (RAM and CPU) is specified in table \ref{tab:specs}, and the computation time of the benchmark reductions for the different systems is plotted in figure \ref{fig:barplot}.

\begin{table}[h]
    \centering
    \begin{tabular}{lcc}
        \toprule
        \textbf{System} & \textbf{RAM} & \textbf{CPU} \\
        \midrule
        A & DDR5 ($2 \times 12$ GB) & Intel i7-13650HX \\
        B & DDR5 ($2 \times 16$ GB)  & AMD Ryzen 9 7950X \\
        C & DDR5 ($4 \times 32$ GB)  & AMD Ryzen 9 9950X \\
        \bottomrule
    \end{tabular}
    \caption{Relevant hardware specifications of the three systems used as benchmark.}
    \label{tab:specs}
\end{table}

The first test (BM1) is to reproduce the factorization of arbitrary 2-loop bosonic sum-integrals, comparing against the formula given in section \ref{sec:factorization}. We find full agreement in an input set of 50 bosonic and 50 fermionic sum-integrals with $-2 < r < 8$ and $0 < s < 4$.

Secondly, we cross-check the 3-loop scalar mass in the DR of the Abelian Higgs model, as recently computed in \cite{Bernardo:2026whs} (BM2). To do this, we employ a routine based on \texttt{FeynRules} \cite{Christensen:2008py} + \texttt{FeynArts} \cite{Hahn:2000kx} + \texttt{FeynCalc} \cite{Shtabovenko:2023idz} to carry out the off-shell matching to the corresponding 3D EFT and extract the expression of the 3-loop scalar mass as a function of a series of unknown 3-loop bosonic sum-integrals. We feed these expressions to \texttt{SIRENA} and reproduce the results in \cite{Bernardo:2026whs} in the same basis of MSI (by setting them as \texttt{priority}).

In order to further demonstrate the potential of \texttt{SIRENA} in state-of-the-art, high-precision computations in thermal QFT, we also provide for the first time the reduction of a new set of 3-loop fermionic sum-integrals appearing in the DR of an Abelian Higgs model coupled to fermions, defined by the 4D Euclidean Lagrangian
\begin{align}
\label{eq:lag4d}
\mathcal{L} &=
    \frac{1}{4} F_{\mu\nu} F_{\mu\nu}
  + (D_\mu \phi)^\dagger (D_\mu \phi)
  + \mu^2 \phi^\dagger \phi
  + \lambda (\phi^\dagger \phi)^2
  \nonumber \\ 
  &+ \bar{\psi} \slashed{D} \psi + y (\phi \bar{\psi}_L \psi_R + \text{h.c.})+ \frac{1}{2 \xi} \left( \partial_\mu B_\mu \right)^2
  \,,
\end{align}
where
$F_{\mu\nu} = \partial_\mu B_\nu - \partial_\nu B_\mu$
is the field strength tensor and
the covariant derivative is defined as
$D_\mu = \partial_\mu - g B_\mu$,
with
$g$ being the gauge coupling. Both $\phi$ and $\psi_L$ have unit charge under $\mathrm{U}(1)$, and $\psi_R$ is taken to be neutral. We assume the following power counting in terms of the gauge coupling $m^2/T^2 \sim |\mathbf{p}|^2/T^2 \sim \lambda \sim y^2 \sim g^2$ and work to $\mathcal{O}(g^6)$.

We focus on the 3-loop scalar mass $m_3^2$, specifically on its $\xi$-dependent terms (BM3), as all sum-integrals appearing in them must necessarily factorize to products of 1-loop ones. The reason is that, via field redefinitions in the 3D EFT \cite{Bernardo:2025vkz,Bernardo:2026whs}, the $\xi$-dependence of $m_3^2$ is known to cancel against products of 1- and 2-loop contributions from $m_3^2$ and a redundant dimension-6 operator, and all of these factorize (see section \ref{sec:factorization}). In this IBP reduction we identify a series of new relations between fermionic 3-loop sum-integrals of mass dimension 2. As a particular example, we have
\begin{equation}
    I_{111110;001} = \frac{8}{(d-3)^2} I_{220011;101}^{110} + \frac{1}{d-4} \left( I_{210011;011} + I_{210011;111} \right) \,,
\end{equation}
which relates a spectacles-type sum-integral with a combination of basketballs. Given the divergent prefactor appearing in this reduction, it is more convenient to express $I_{220011;101}$ in terms of the others, avoiding the need of taking its evaluation in an $\epsilon$-expansion to high orders. In this and other examples in the reduction, we observe the proliferation of such divergent prefactors when reducing \textit{more complex} topologies to \textit{simpler} basketballs. The study of singularity-free MSI bases in the 3-loop fermionic sector constitutes a promising future line of work, in line with recently developed techniques in the field of regular Feynman integrals \cite{DeAngelis:2025agn}.

Back to the original purpose, the combined reduction result in the $\xi$-dependent part of $m_3^2$ is such that all 3-loop MSI cancel, and it is indeed expressed as a combination of factorized 1-loop cases, \textit{viz.}
\begin{align}
    m_3^2 \big|_{\xi} &= \left[ \left( (2 - 2d)g^6 - 4 d g^4 \lambda - 16 g^2 \lambda^2  \right) \xi + \left( \frac{d}{2} g^6 + 2 g^4 \lambda \right) \xi^2 \right] I_{2} I_{2} I_{1} \nonumber \\
    &+ \left[ 4 \left( g^6 + 2 g^2 y^2 \lambda \right) \xi - g^4 y^2 \xi^2 \right] I_{2} I_{2} I_{1;1} + 2 \left( g^4 y^2 + g^2 y^4 \right) \xi I_{2} I_{2;1} I_{1} \nonumber \\
    &- 2 \left( g^4 y^2 + g^2 y^4 \right) \xi I_{2;1} I_{1} I_{1;1} + \left(-d^2 g^6 - 8 d g^4 \lambda - 16 g^2 \lambda^2 \right) \xi I_{3} I_{1} I_{1} \nonumber \\
    &+ 4 \left(d g^4 y^2 + 4 g^2 y^2 \lambda \right) \xi I_{3} I_{1} I_{1;1} - 4 g^2 y^4 \xi I_{3} I_{1;1} I_{1;1} \,,
\end{align}
where all fermionic 1-loop sum-integrals may be further simplified to their bosonic counterparts via even-odd splitting (see eq. \eqref{eq:even-odd 1-loop}).

In figure \ref{fig:barplot} we observe that benchmarks involving fermionic sum-integrals are in general more computationally expensive. This is natural, as the existence of different possible signatures (instead of the single zero signature in fully bosonic reductions) enlarge the set of independent seeds and thus the typical size of the IBP systems to solve.

While \texttt{SIRENA} is not hard-coded for 2- and 3-loop sum-integrals, the performance of the code has only been tested on these cases, as sum-integrals beyond 3-loop order have rarely been explored in the literature (see~\cite{Navarrete:2024ruu} for an exception). Performance checks on a few 4-loop sum-integral examples, which notably increase computation times, suggest the need to employ alternative techniques that avoid dealing with large sets of seeds (see~\textit{e.g.} \cite{Smirnov:2025dfy} for a thorough review on different approaches). We leave the exploration of these methods in the context of sum-integral reduction to future work.

\section{Conclusions}
\label{sec:conclusions}

In this work we presented \texttt{SIRENA}, the first public implementation of the Laporta algorithm for the reduction of multi-loop sum-integrals to masters. Most of the techniques and algorithms employed are extensions of well-established methods for standard Feynman integrals, adapted to consistently account for the Matsubara sum structure characteristic of finite-temperature QFT. The package is designed to minimize user intervention, allowing it to be applied straightforwardly to the lengthy expressions that commonly arise in high-order perturbative computations. 

As an additional result, building upon previous work \cite{Davydychev:2023jto}, we have derived an algebraic formula for the factorization of fermionic 2-loop sum-integrals to products of 1-loop masters. This result fully removes non-factorizable 2-loop sum-integrals from the perturbative expansions arising in DR, and constitutes a crucial step towards the full automation of finite temperature matching up to 2-loop order in a future version of \texttt{Matchotter} \cite{Fuentes-Martin:2026bhr}. For higher orders, similar factorization formulae are not known to exist, so automatic IBP reduction remains an essential tool.

To demonstrate the capabilities of \texttt{SIRENA}, we have reproduced several non-trivial results from the literature up to 3-loop order and, for the first time, provided reductions for some 3-loop fermionic sum-integrals appearing in a gauge theory with fermions. While the code has only been fully tested at 2- and 3-loop order, it is in principle capable of reducing sum-integrals at higher loop orders as well. We therefore expect this tool to facilitate broader access to high-order perturbative calculations in finite temperature quantum field theory, at a time when precision plays an increasingly central role.

\section*{Acknowledgments}
We thank Mikael Chala for his help and invaluable support in all aspects of this work. LG is indebted to Pablo Navarrete for useful discussions. LG, JLM and AMS are supported by the FPU program under grants number FPU23/02026, FPU23/02028 and FPU23/01639, respectively. AMS is further supported by grants CNS2024-154834 and EUR2024.153549 by the Spanish Research Agency  and the European Union (NextGenerationEU/PRTR and FEDER/UE). This work has received funding from the European Research Council under grant agreement n. 101230200, MICIU/AEI/10.13039/501100011033 and ERDF/EU (grant PID2022-139466NB-C22) as well as from Junta de Andaluc\'ia (grant P21-00199).

\appendix

\section{Code snippet of 2-loop factorization formula}
\label{app:snippet}
In this appendix we provide a \texttt{Mathematica} code snippet that computes the factorization of any 2-loop sum-integral of arbitrary signature, based on \cite{Davydychev:2023jto} and our formula in eq.~\eqref{eq: Final Fact Formula}. 

The function \texttt{Int} thus takes two arguments: (1) the set of $\alpha_i$ and $\beta_i$ given as a list, and (2) a list \texttt{\{s1,s2\}}, which specifies the signature of the two loop momenta. Mixed signature sum-integrals are factorized by first employing the relations in eq. \eqref{eq: relations_mixed_momenta} and then the formula is applied to the result. The output is given as a combination of bosonic \texttt{iB[a,b]} and fermionic \texttt{iF[a,b]} 1-loop sum-integrals, with their two arguments denoting their propagator and numerator powers, respectively.

This code can be used as a standalone reduction tool at 2-loops, but it also served as a thorough cross-check of the IBP-based results produced by \texttt{SIRENA} (see section~\ref{sec:examples}).

\newpage
\begin{lstlisting}[style=mathematica, basicstyle=\fontsize{9.5}{11}\selectfont\ttfamily]
c[asum_, a1_, a2_, j_] := Module[{nj = Ceiling[(asum + j)/2], a3 = asum - a1 - a2},
    (-1)^(asum - nj + 1)  Pochhammer[1 - d/2, nj - j - 1]/  
        (2  Pochhammer[1/2, nj - a2 -a3] Pochhammer[1/2, nj - j - a1 - a3]  (a3 - 1)!)  
    Sum[Pochhammer[d/2 - nj + 1, k - 1] (nj - k - 1)!/
        (Pochhammer[d/2 + 3/2 - asum, nj - k]  Pochhammer[1/2, a3 - nj + k] 
        (k - 1)!  (k - j - 1)!  (a2 - k)!  (a1 - k + j)!), 
    {k, Max[1 + j, 1], Min[a1 + j, a2]}]];
    
h[n_, type_] := Module[{s = Max[n, 1]}, 
   Gamma[s] *(Switch[type, 0, iB, 1, iF][s, 2 s - 2 n])/2/ Pochhammer[1 - d/2, s - 1]];
   
Int[{a1_, a2_, a3_, b1_, b2_, b3_}, {s1_, s2_}] := Switch[{s1, s2}, {1, 0}, 
    Int[{a1, a3, a2, b1, b3, b2}, {1, 1}], {0, 1}, 
    (-1)^b1  Int[{a3, a2, a1, b3, b2, b1}, {1, 1}], {_, _},
   Which[a1 < a2, (-1)^b3  Int[{a2, a1, a3, b2, b1, b3}, {s1, s2}], 
   (a2 < a3) && ({s1, s2} === {0, 0}), Int[{a1, a3, a2, b1, b3, b2}, {s1, s2}], True, 
    Collect[Module[{asum = a1 + a2 + a3, bsum = b1 + b2 + b3}, (1 + (-1)^bsum)*
        If[a1*a2*a3 > 0, (-1)^asum*(Sum[c[asum, a1, a2, j]  Sum[With[{el = asum - b1 - j - k}, 
            Binomial[b3, k]  (-1)^b2  (1 + (-1)^el)  
            h[el/2, s1]  h[asum - bsum/2 - el/2, s1]], {k, 0, b3}], {j, 1 - a1, a2 - 1}] + 
        Sum[c[asum, a1, a3, j]  Sum[With[{el = asum - b1 - j - k}, 
            Binomial[b2, k]  (-1)^b3  (1 + (-1)^el)  
            h[el/2, s1]  h[asum - bsum/2 - el/2, 0]], {k, 0, b2}], {j, 1 - a1, a3 - 1}] + 
        Sum[c[asum, a2, a3, j]  Sum[With[{el = asum - b2 - j - k}, 
            Binomial[b1, k]  (1 + (-1)^el)  
            h[el/2, s1]  h[asum - bsum/2 - el/2, 0]], {k, 0, b1}], {j, 1 - a2, a3 - 1}]), 
        If[a3 < 0, 2  (-1)^b2  (-a3)!  Sum[With[{x = Mod[b2 + n, 2]}, 
        Sum[With[{y = k - j - Floor[(b2 + n)/2], z = asum + n - k}, 
            Binomial[b3, 2  j - x]  2^n/(l!  (n - 2  l)!  (a2 - k)!  (a1 - z)!  
            Gamma[z]  Gamma[k]  4^l)  Pochhammer[d/2, l]  Pochhammer[1 - d/2, z - 1]  
            Pochhammer[1 - d/2, k - 1]/(Pochhammer[d/2 + 1 - z, l]  Pochhammer[d/2 + 1 - k, l])
            h[asum - bsum/2 - y, s1]  h[y, s1]], {j, x, Floor[(b3 + x)/2]}, 
        {k, Max[1, n + asum - a1], Min[a2, n + asum - 1]}, {l, 0, Floor[n/2]}]], {n, 0, -a3}], 
         2  (-1)^b3  (-a2)!  Sum[With[{x = Mod[b3 + n, 2]}, 
        Sum[With[{y = k - j - Floor[(b3 + n)/2], z = asum + n - k}, 
            Binomial[b2, 2  j - x]  2^n/(l!  (n - 2  l)!  (a3 - k)!  (a1 - z)!  
            Gamma[z]  Gamma[k]  4^l)  Pochhammer[d/2, l]  Pochhammer[1 - d/2, z - 1]  
            Pochhammer[1 - d/2, k - 1]/(Pochhammer[d/2 + 1 - z, l]  Pochhammer[d/2 + 1 - k, l])  
            h[asum - bsum/2 - y, s1]  h[y, 0]], {j, x, Floor[(b2 + x)/2]}, 
        {k, Max[1, n + asum - a1], Min[a3, n + asum - 1]}, {l, 0, Floor[n/2]}]], {n, 0, -a2}]
    ]]],{(iB | iF)[__]}, Factor]]];
\end{lstlisting}

\bibliographystyle{style} 

\bibliography{refs} 

@article{Matsubara:1955ws,
    author = "Matsubara, Takeo",
    title = "{A New approach to quantum statistical mechanics}",
    doi = "10.1143/PTP.14.351",
    journal = "Prog. Theor. Phys.",
    volume = "14",
    pages = "351--378",
    year = "1955"
}

@article{Ginsparg:1980ef,
    author = "Ginsparg, Paul H.",
    title = "{First Order and Second Order Phase Transitions in Gauge Theories at Finite Temperature}",
    reportNumber = "SACLAY-DPh-T 80/27",
    doi = "10.1016/0550-3213(80)90418-6",
    journal = "Nucl. Phys. B",
    volume = "170",
    pages = "388--408",
    year = "1980"
}

@article{Appelquist:1981vg,
    author = "Appelquist, Thomas and Pisarski, Robert D.",
    title = "{High-Temperature Yang-Mills Theories and Three-Dimensional Quantum Chromodynamics}",
    reportNumber = "Print-81-0020 (YALE), YTP-81-01, COO-3075-203",
    doi = "10.1103/PhysRevD.23.2305",
    journal = "Phys. Rev. D",
    volume = "23",
    pages = "2305",
    year = "1981"
}

@article{Ekstedt:2022bff,
    author = "Ekstedt, Andreas and Schicho, Philipp and Tenkanen, Tuomas V. I.",
    title = "{DRalgo: A package for effective field theory approach for thermal phase transitions}",
    eprint = "2205.08815",
    archivePrefix = "arXiv",
    primaryClass = "hep-ph",
    reportNumber = "HIP-2022-11/TH, NORDITA 2022-030",
    doi = "10.1016/j.cpc.2023.108725",
    journal = "Comput. Phys. Commun.",
    volume = "288",
    pages = "108725",
    year = "2023"
}

@article{Kajantie:1996mn,
    author = "Kajantie, K. and Laine, M. and Rummukainen, K. and Shaposhnikov, Mikhail E.",
    title = "{Is there a~ hot electroweak phase transition at $m_H \gtrsim m_W$?}",
    eprint = "hep-ph/9605288",
    archivePrefix = "arXiv",
    reportNumber = "CERN-TH-96-126, HD-THEP-96-15, IUHET-333",
    doi = "10.1103/PhysRevLett.77.2887",
    journal = "Phys. Rev. Lett.",
    volume = "77",
    pages = "2887--2890",
    year = "1996"
}

@article{Gurtler:1997hr,
    author = "Gurtler, M. and Ilgenfritz, Ernst-Michael and Schiller, A.",
    title = "{Where the electroweak phase transition ends}",
    eprint = "hep-lat/9704013",
    archivePrefix = "arXiv",
    reportNumber = "UL-NTZ-10-97, HUB-EP-97-24, DESY-97-086",
    doi = "10.1103/PhysRevD.56.3888",
    journal = "Phys. Rev. D",
    volume = "56",
    pages = "3888--3895",
    year = "1997"
}

@article{Csikor:1998eu,
    author = "Csikor, F. and Fodor, Z. and Heitger, J.",
    title = "{Endpoint of the hot electroweak phase transition}",
    eprint = "hep-ph/9809291",
    archivePrefix = "arXiv",
    reportNumber = "ITP-BUDAPEST-541, KEK-TH-580, KEK-PREPRINT-98-160, MS-TPI-98-16",
    doi = "10.1103/PhysRevLett.82.21",
    journal = "Phys. Rev. Lett.",
    volume = "82",
    pages = "21--24",
    year = "1999"
}

@article{Gould:2023ovu,
    author = "Gould, Oliver and Tenkanen, Tuomas V. I.",
    title = "{Perturbative effective field theory expansions for cosmological phase transitions}",
    eprint = "2309.01672",
    archivePrefix = "arXiv",
    primaryClass = "hep-ph",
    reportNumber = "NORDITA 2023-037",
    doi = "10.1007/JHEP01(2024)048",
    journal = "JHEP",
    volume = "01",
    pages = "048",
    year = "2024"
}

@article{Niemi:2020hto,
    author = "Niemi, Lauri and Ramsey-Musolf, Michael J. and Tenkanen, Tuomas V. I. and Weir, David J.",
    title = "{Thermodynamics of a Two-Step Electroweak Phase Transition}",
    eprint = "2005.11332",
    archivePrefix = "arXiv",
    primaryClass = "hep-ph",
    reportNumber = "HIP-2020-11/TH, ACFI-T20-05",
    doi = "10.1103/PhysRevLett.126.171802",
    journal = "Phys. Rev. Lett.",
    volume = "126",
    number = "17",
    pages = "171802",
    year = "2021"
}

@article{Niemi:2022bjg,
    author = {Niemi, Lauri and Rummukainen, Kari and Sepp{\"a}, Riikka and Weir, David J.},
    title = "{Infrared physics of the 3D SU(2) adjoint Higgs model at the crossover transition}",
    eprint = "2206.14487",
    archivePrefix = "arXiv",
    primaryClass = "hep-lat",
    reportNumber = "HIP-2022-18/TH",
    doi = "10.1007/JHEP02(2023)212",
    journal = "JHEP",
    volume = "02",
    pages = "212",
    year = "2023"
}

@article{Gould:2024chm,
    author = "Gould, Oliver and Kormu, Anna and Weir, David J.",
    title = "{Nonperturbative test of nucleation calculations for strong phase transitions}",
    eprint = "2404.01876",
    archivePrefix = "arXiv",
    primaryClass = "hep-th",
    reportNumber = "HIP-2024-9/TH",
    doi = "10.1103/PhysRevD.111.L051901",
    journal = "Phys. Rev. D",
    volume = "111",
    number = "5",
    pages = "L051901",
    year = "2025"
}

@article{Chala:2025oul,
    author = "Chala, Mikael and Gil, Luis and Ren, Zhe",
    title = "{Phase transitions in dimensional reduction up to three loops}",
    eprint = "2505.14335",
    archivePrefix = "arXiv",
    primaryClass = "hep-ph",
    doi = "10.1088/1674-1137/adf322",
    journal = "Chin. Phys.",
    volume = "49",
    number = "12",
    pages = "123105",
    year = "2025"
}

@article{Bernardo:2026whs,
    author = "Bernardo, Fabio and Chala, Mikael and Gil, Luis and Schicho, Philipp",
    title = "{Hard thermal contributions to phase transition observables at NNLO}",
    eprint = "2602.06962",
    archivePrefix = "arXiv",
    primaryClass = "hep-ph",
    month = "2",
    year = "2026"
}

@article{Fuentes-Martin:2026bhr,
    author = "Fuentes-Mart{\'\i}n, Javier and L{\'o}pez Miras, Javier and Moreno-S{\'a}nchez, Adri{\'a}n",
    title = "{Matchotter: An Automated Tool for Dimensional Reduction at Finite Temperature}",
    eprint = "2604.21972",
    archivePrefix = "arXiv",
    primaryClass = "hep-ph",
    month = "4",
    year = "2026"
}

@phdthesis{Schicho:2020xaf,
    author = "Schicho, Philipp",
    title = "{Multi-loop investigations of strong interactions at high temperatures}",
    doi = "10.24442/BORISTHESES.1988",
    school = "U. Bern",
    year = "2020"
}

@article{Andersen:2008bz,
    author = "Andersen, Jens O. and Kyllingstad, Lars",
    title = "{Four-loop Screened Perturbation Theory}",
    eprint = "0805.4478",
    archivePrefix = "arXiv",
    primaryClass = "hep-ph",
    doi = "10.1103/PhysRevD.78.076008",
    journal = "Phys. Rev. D",
    volume = "78",
    pages = "076008",
    year = "2008"
}

@article{Gynther:2007bw,
    author = "Gynther, A. and Laine, M. and Schr{\"o}der, Y. and Torrero, C. and Vuorinen, A.",
    title = "{Four-loop pressure of massless O(N) scalar field theory}",
    eprint = "hep-ph/0703307",
    archivePrefix = "arXiv",
    reportNumber = "BI-TP-2007-05, ECT-07-06",
    doi = "10.1088/1126-6708/2007/04/094",
    journal = "JHEP",
    volume = "04",
    pages = "094",
    year = "2007"
}

@article{Navarrete:2025yxy,
    author = {Navarrete, Pablo and Paatelainen, Risto and Sepp{\"a}nen, Kaapo and Tenkanen, Tuomas V. I.},
    title = "{Cosmological phase transitions without high-temperature expansions}",
    eprint = "2507.07014",
    archivePrefix = "arXiv",
    primaryClass = "hep-ph",
    reportNumber = "HIP-2025-20/TH",
    doi = "10.1007/JHEP01(2026)113",
    journal = "JHEP",
    volume = "01",
    pages = "113",
    year = "2026"
}

@phdthesis{Seppanen:2025owq,
    author = {Sepp{\"a}nen, Kaapo},
    title = "{Quark Matter Thermodynamics from High-Order Perturbative QCD}",
    school = "Helsinki U.",
    year = "2025"
}

@article{Ghisoiu:2012kn,
    author = "Ghisoiu, Ioan and Schr{\"o}der, York",
    title = "{A new three-loop sum-integral of mass dimension two}",
    eprint = "1207.6214",
    archivePrefix = "arXiv",
    primaryClass = "hep-ph",
    reportNumber = "BI-TP-2012-30",
    doi = "10.1007/JHEP09(2012)016",
    journal = "JHEP",
    volume = "09",
    pages = "016",
    year = "2012"
}

@phdthesis{Ghisoiu:2013zoj,
    author = "Ghisoiu, Ioan",
    title = "{Three-loop Debye mass and effective coupling in thermal QCD}",
    school = "U. Bielefeld (main)",
    year = "2013"
}

@article{Arnold:1994eb,
    author = "Arnold, Peter Brockway and Zhai, Cheng-xing",
    title = "{The Three loop free energy for high temperature QED and QCD with fermions}",
    eprint = "hep-ph/9410360",
    archivePrefix = "arXiv",
    reportNumber = "UW-PT-94-11",
    doi = "10.1103/PhysRevD.51.1906",
    journal = "Phys. Rev. D",
    volume = "51",
    pages = "1906--1918",
    year = "1995"
}

@article{Nishimura:2012ee,
    author = "Nishimura, M. and Schr{\"o}der, Y.",
    title = "{IBP methods at finite temperature}",
    eprint = "1207.4042",
    archivePrefix = "arXiv",
    primaryClass = "hep-ph",
    reportNumber = "BI-TP-2012-28",
    doi = "10.1007/JHEP09(2012)051",
    journal = "JHEP",
    volume = "09",
    pages = "051",
    year = "2012"
}

@article{Ghisoiu:2012yk,
    author = "Ghisoiu, Ioan and Schr{\"o}der, York",
    title = "{A New Method for Taming Tensor Sum-Integrals}",
    eprint = "1208.0284",
    archivePrefix = "arXiv",
    primaryClass = "hep-ph",
    reportNumber = "BI-TP-2012-32",
    doi = "10.1007/JHEP11(2012)010",
    journal = "JHEP",
    volume = "11",
    pages = "010",
    year = "2012"
}

@article{Navarrete:2022adz,
    author = {Navarrete, Pablo and Schr{\"o}der, York},
    title = "{Tackling the infamous $g^6$ term of the QCD pressure}",
    eprint = "2207.10151",
    archivePrefix = "arXiv",
    primaryClass = "hep-ph",
    doi = "10.22323/1.416.0014",
    journal = "PoS",
    volume = "LL2022",
    pages = "014",
    year = "2022"
}

@article{Navarrete:2024ruu,
    author = {Navarrete, Pablo and Schr{\"o}der, York},
    title = "{The g$^{6}$ pressure of hot Yang-Mills theory: canonical form of the integrand}",
    eprint = "2408.15830",
    archivePrefix = "arXiv",
    primaryClass = "hep-ph",
    doi = "10.1007/JHEP11(2024)037",
    journal = "JHEP",
    volume = "11",
    pages = "037",
    year = "2024"
}

@article{Moller:2010xw,
    author = "Moller, Jan and Schr{\"o}der, York",
    editor = {Bl{\"u}mlein, Johannes and Moch, Sven-Olaf and Riemann, Tord},
    title = "{Open problems in hot QCD}",
    eprint = "1007.1223",
    archivePrefix = "arXiv",
    primaryClass = "hep-ph",
    reportNumber = "BI-TP-2010-20",
    doi = "10.1016/j.nuclphysbps.2010.08.046",
    journal = "Nucl. Phys. B Proc. Suppl.",
    volume = "205-206",
    pages = "218--223",
    year = "2010"
}

@article{Moller:2012chx,
    author = "Moller, J. and Schr{\"o}der, Y.",
    title = "{Three-loop matching coefficients for hot QCD: Reduction and gauge independence}",
    eprint = "1207.1309",
    archivePrefix = "arXiv",
    primaryClass = "hep-ph",
    reportNumber = "BI-TP-2012-25",
    doi = "10.1007/JHEP08(2012)025",
    journal = "JHEP",
    volume = "08",
    pages = "025",
    year = "2012"
}

@article{Arnold:1994ps,
    author = "Arnold, Peter Brockway and Zhai, Cheng-Xing",
    title = "{The Three loop free energy for pure gauge QCD}",
    eprint = "hep-ph/9408276",
    archivePrefix = "arXiv",
    reportNumber = "UW-PT-94-03",
    doi = "10.1103/PhysRevD.50.7603",
    journal = "Phys. Rev. D",
    volume = "50",
    pages = "7603--7623",
    year = "1994"
}

@article{Schroder:2012hm,
    author = "Schr{\"o}der, York",
    title = "{A fresh look on three-loop sum-integrals}",
    eprint = "1207.5666",
    archivePrefix = "arXiv",
    primaryClass = "hep-ph",
    reportNumber = "BI-TP-2012-29",
    doi = "10.1007/JHEP08(2012)095",
    journal = "JHEP",
    volume = "08",
    pages = "095",
    year = "2012"
}

@article{Davydychev:2023jto,
    author = {Davydychev, Andrei I. and Navarrete, Pablo and Schr{\"o}der, York},
    title = "{Factorizing two-loop vacuum sum-integrals}",
    eprint = "2312.17367",
    archivePrefix = "arXiv",
    primaryClass = "hep-ph",
    doi = "10.1007/JHEP02(2024)104",
    journal = "JHEP",
    volume = "02",
    pages = "104",
    year = "2024"
}

@article{Davydychev:2022dcw,
    author = {Davydychev, Andrei I. and Schr{\"o}der, York},
    title = "{Recursion-free solution for two-loop vacuum integrals with {\textquotedblleft}collinear{\textquotedblright} masses}",
    eprint = "2210.10593",
    archivePrefix = "arXiv",
    primaryClass = "hep-ph",
    doi = "10.1007/JHEP12(2022)047",
    journal = "JHEP",
    volume = "12",
    pages = "047",
    year = "2022"
}

@article{Ghisoiu:2015uza,
    author = "Ghisoiu, Ioan and Moller, Jan and Schr{\"o}der, York",
    title = "{Debye screening mass of hot Yang-Mills theory to three-loop order}",
    eprint = "1509.08727",
    archivePrefix = "arXiv",
    primaryClass = "hep-ph",
    doi = "10.1007/JHEP11(2015)121",
    journal = "JHEP",
    volume = "11",
    pages = "121",
    year = "2015"
}

@article{Arnold:1992rz,
    author = "Arnold, Peter Brockway and Espinosa, Olivier",
    title = "{The Effective potential and first order phase transitions: Beyond leading-order}",
    eprint = "hep-ph/9212235",
    archivePrefix = "arXiv",
    reportNumber = "UW-PT-92-18, USM-TH-60",
    doi = "10.1103/PhysRevD.47.3546",
    journal = "Phys. Rev. D",
    volume = "47",
    pages = "3546",
    year = "1993",
    note = "[Erratum: Phys.Rev.D 50, 6662 (1994)]"
}

@article{Coriano:1994re,
    author = "Coriano, Claudio and Parwani, Rajesh R.",
    title = "{The Three loop equation of state of QED at high temperature}",
    eprint = "hep-ph/9405343",
    archivePrefix = "arXiv",
    reportNumber = "ANL-HEP-PR-94-02, SACLAY-SPH-T-94-054",
    doi = "10.1103/PhysRevLett.73.2398",
    journal = "Phys. Rev. Lett.",
    volume = "73",
    pages = "2398--2401",
    year = "1994"
}

@article{Tkachov:1981wb,
    author = "Tkachov, F. V.",
    title = "{A theorem on analytical calculability of 4-loop renormalization group functions}",
    doi = "10.1016/0370-2693(81)90288-4",
    journal = "Phys. Lett. B",
    volume = "100",
    pages = "65--68",
    year = "1981"
}

@article{Chetyrkin:1981qh,
    author = "Chetyrkin, K. G. and Tkachov, F. V.",
    title = "{Integration by parts: The algorithm to calculate $\beta$-functions in 4 loops}",
    doi = "10.1016/0550-3213(81)90199-1",
    journal = "Nucl. Phys. B",
    volume = "192",
    pages = "159--204",
    year = "1981"
}

@article{Beneke:1997zp,
    author = "Beneke, M. and Smirnov, Vladimir A.",
    title = "{Asymptotic expansion of Feynman integrals near threshold}",
    eprint = "hep-ph/9711391",
    archivePrefix = "arXiv",
    reportNumber = "CERN-TH-97-315",
    doi = "10.1016/S0550-3213(98)00138-2",
    journal = "Nucl. Phys. B",
    volume = "522",
    pages = "321--344",
    year = "1998"
}

@article{Laporta:2000dsw,
    author = "Laporta, S.",
    title = "{High-precision calculation of multiloop Feynman integrals by difference equations}",
    eprint = "hep-ph/0102033",
    archivePrefix = "arXiv",
    doi = "10.1142/S0217751X00002159",
    journal = "Int. J. Mod. Phys. A",
    volume = "15",
    pages = "5087--5159",
    year = "2000"
}

@article{Smirnov:2025dfy,
    author = "Smirnov, Alexander and Smirnov, Vladimir",
    title = "{Two decades of algorithmic Feynman integral reduction}",
    eprint = "2510.10748",
    archivePrefix = "arXiv",
    primaryClass = "hep-th",
    month = "10",
    year = "2025"
}

@article{Anastasiou:2004vj,
    author = "Anastasiou, Charalampos and Lazopoulos, Achilleas",
    title = "{Automatic integral reduction for higher order perturbative calculations}",
    eprint = "hep-ph/0404258",
    archivePrefix = "arXiv",
    reportNumber = "SLAC-PUB-10428",
    doi = "10.1088/1126-6708/2004/07/046",
    journal = "JHEP",
    volume = "07",
    pages = "046",
    year = "2004"
}

@article{Smirnov:2025prc,
    author = "Smirnov, Alexander V. and Zeng, Mao",
    title = "{FIRE 7: Automatic Reduction with Modular Approach}",
    eprint = "2510.07150",
    archivePrefix = "arXiv",
    primaryClass = "hep-ph",
    month = "10",
    year = "2025"
}

@article{vonManteuffel:2012np,
    author = "von Manteuffel, A. and Studerus, C.",
    title = "{Reduze 2 - Distributed Feynman Integral Reduction}",
    eprint = "1201.4330",
    archivePrefix = "arXiv",
    primaryClass = "hep-ph",
    reportNumber = "ZU-TH-01-12, MZ-TH-12-03, BI-TP-2012-02",
    month = "1",
    year = "2012"
}

@article{Lee:2012cn,
    author = "Lee, R. N.",
    title = "{Presenting LiteRed: a tool for the Loop InTEgrals REDuction}",
    eprint = "1212.2685",
    archivePrefix = "arXiv",
    primaryClass = "hep-ph",
    month = "12",
    year = "2012"
}

@article{Maierhofer:2017gsa,
    author = {Maierh{\"o}fer, Philipp and Usovitsch, Johann and Uwer, Peter},
    title = "{Kira{\textemdash}A Feynman integral reduction program}",
    eprint = "1705.05610",
    archivePrefix = "arXiv",
    primaryClass = "hep-ph",
    doi = "10.1016/j.cpc.2018.04.012",
    journal = "Comput. Phys. Commun.",
    volume = "230",
    pages = "99--112",
    year = "2018"
}

@article{Lange:2025fba,
    author = "Lange, Fabian and Usovitsch, Johann and Wu, Zihao",
    title = "{Kira 3: integral reduction with efficient seeding and optimized equation selection}",
    eprint = "2505.20197",
    archivePrefix = "arXiv",
    primaryClass = "hep-ph",
    reportNumber = "ZU-TH 39/25, HU-EP-25/17-RTG",
    doi = "10.1016/j.cpc.2025.109999",
    journal = "Comput. Phys. Commun.",
    volume = "322",
    pages = "109999",
    year = "2026"
}

@article{Wu:2023upw,
    author = "Wu, Zihao and Boehm, Janko and Ma, Rourou and Xu, Hefeng and Zhang, Yang",
    title = "{NeatIBP 1.0, a package generating small-size integration-by-parts relations for Feynman integrals}",
    eprint = "2305.08783",
    archivePrefix = "arXiv",
    primaryClass = "hep-ph",
    reportNumber = "USTC-ICTS/PCFT-23-15",
    doi = "10.1016/j.cpc.2023.108999",
    journal = "Comput. Phys. Commun.",
    volume = "295",
    pages = "108999",
    year = "2024"
}

@article{Shih:2026jfe,
    author = "Shih, David",
    title = "{Learning to Unscramble Feynman Loop Integrals with SAILIR}",
    eprint = "2604.05034",
    archivePrefix = "arXiv",
    primaryClass = "hep-ph",
    journal = "",
    month = "4",
    year = "2026"
}

@manual{flint,
  key = {{FLINT}},
  author = {{The {FLINT} team}},
  title = {{FLINT}: {F}ast {L}ibrary for {N}umber {T}heory},
  year = {2025},
  note = {Version 3.4.0, \url{https://flintlib.org}}
}

@article{sympy,
 title = {SymPy: symbolic computing in Python},
 author = {Meurer, Aaron and Smith, Christopher P. and Paprocki, Mateusz and \v{C}ert\'{i}k, Ond\v{r}ej and Kirpichev, Sergey B. and Rocklin, Matthew and Kumar, AMiT and Ivanov, Sergiu and Moore, Jason K. and Singh, Sartaj and Rathnayake, Thilina and Vig, Sean and Granger, Brian E. and Muller, Richard P. and Bonazzi, Francesco and Gupta, Harsh and Vats, Shivam and Johansson, Fredrik and Pedregosa, Fabian and Curry, Matthew J. and Terrel, Andy R. and Rou\v{c}ka, \v{S}t\v{e}p\'{a}n and Saboo, Ashutosh and Fernando, Isuru and Kulal, Sumith and Cimrman, Robert and Scopatz, Anthony},
 year = 2017,
 month = jan,
 volume = 3,
 pages = {e103},
 journal = {PeerJ Computer Science},
 issn = {2376-5992},
 url = {https://doi.org/10.7717/peerj-cs.103},
 doi = {10.7717/peerj-cs.103}
}

@misc{pybind11,
   author = "{W. Jakob, J. Rhinelander, D. Moldovan and \href{https://github.com/pybind/pybind11/graphs/contributors}{others}}",
   year = {2017},
   title = "{\href{https://github.com/pybind/pybind11}{pybind11 -- Seamless operability between C++11 and Python}}"
}

@article{Christensen:2008py,
    author = "Christensen, Neil D. and Duhr, Claude",
    title = "{FeynRules - Feynman rules made easy}",
    eprint = "0806.4194",
    archivePrefix = "arXiv",
    primaryClass = "hep-ph",
    reportNumber = "MSUHEP-080616, CP3-08-20",
    doi = "10.1016/j.cpc.2009.02.018",
    journal = "Comput. Phys. Commun.",
    volume = "180",
    pages = "1614--1641",
    year = "2009"
}

@article{Hahn:2000kx,
    author = "Hahn, Thomas",
    title = "{Generating Feynman diagrams and amplitudes with FeynArts 3}",
    eprint = "hep-ph/0012260",
    archivePrefix = "arXiv",
    reportNumber = "KA-TP-23-2000",
    doi = "10.1016/S0010-4655(01)00290-9",
    journal = "Comput. Phys. Commun.",
    volume = "140",
    pages = "418--431",
    year = "2001"
}

@article{Shtabovenko:2023idz,
    author = "Shtabovenko, Vladyslav and Mertig, Rolf and Orellana, Frederik",
    title = "{FeynCalc 10: Do multiloop integrals dream of computer codes?}",
    eprint = "2312.14089",
    archivePrefix = "arXiv",
    primaryClass = "hep-ph",
    reportNumber = "P3H-23-089, TTP23-056, SI-HEP-2023-27",
    doi = "10.1016/j.cpc.2024.109357",
    journal = "Comput. Phys. Commun.",
    volume = "306",
    pages = "109357",
    year = "2025"
}

@article{Schicho:2022wty,
    author = "Schicho, Philipp and Tenkanen, Tuomas V. I. and White, Graham",
    title = "{Combining thermal resummation and gauge invariance for electroweak phase transition}",
    eprint = "2203.04284",
    archivePrefix = "arXiv",
    primaryClass = "hep-ph",
    reportNumber = "HIP-2022-2/TH, NORDITA 2022-009",
    doi = "10.1007/JHEP11(2022)047",
    journal = "JHEP",
    volume = "11",
    pages = "047",
    year = "2022"
}

@article{Gould:2023jbz,
    author = "Gould, Oliver and Xie, Cheng",
    title = "{Higher orders for cosmological phase transitions: a global study in a Yukawa model}",
    eprint = "2310.02308",
    archivePrefix = "arXiv",
    primaryClass = "hep-ph",
    doi = "10.1007/JHEP12(2023)049",
    journal = "JHEP",
    volume = "12",
    pages = "049",
    year = "2023"
}

@article{Kierkla:2023von,
    author = "Kierkla, Maciej and Swiezewska, Bogumila and Tenkanen, Tuomas V. I. and van de Vis, Jorinde",
    title = "{Gravitational waves from supercooled phase transitions: dimensional transmutation meets dimensional reduction}",
    eprint = "2312.12413",
    archivePrefix = "arXiv",
    primaryClass = "hep-ph",
    doi = "10.1007/JHEP02(2024)234",
    journal = "JHEP",
    volume = "02",
    pages = "234",
    year = "2024"
}

@article{Chala:2024xll,
    author = "Chala, Mikael and Criado, Juan Carlos and Gil, Luis and Miras, Javier L{\'o}pez",
    title = "{Higher-order-operator corrections to phase-transition parameters in dimensional reduction}",
    eprint = "2406.02667",
    archivePrefix = "arXiv",
    primaryClass = "hep-ph",
    doi = "10.1007/JHEP10(2024)025",
    journal = "JHEP",
    volume = "10",
    pages = "025",
    year = "2024"
}

@article{Niemi:2024axp,
    author = "Niemi, Lauri and Ramsey-Musolf, Michael J. and Xia, Guotao",
    title = "{Nonperturbative study of the electroweak phase transition in the real scalar singlet extended standard model}",
    eprint = "2405.01191",
    archivePrefix = "arXiv",
    primaryClass = "hep-ph",
    reportNumber = "HIP-2024-7/TH, ACFI T24-03",
    doi = "10.1103/PhysRevD.110.115016",
    journal = "Phys. Rev. D",
    volume = "110",
    number = "11",
    pages = "115016",
    year = "2024"
}

@article{Qin:2024idc,
    author = "Qin, Renhui and Bian, Ligong",
    title = "{First-order electroweak phase transition at finite density}",
    eprint = "2407.01981",
    archivePrefix = "arXiv",
    primaryClass = "hep-ph",
    doi = "10.1007/JHEP08(2024)157",
    journal = "JHEP",
    volume = "08",
    pages = "157",
    year = "2024"
}

@article{Niemi:2024vzw,
    author = "Niemi, Lauri and Tenkanen, Tuomas V. I.",
    title = "{Investigating two-loop effects for first-order electroweak phase transitions}",
    eprint = "2408.15912",
    archivePrefix = "arXiv",
    primaryClass = "hep-ph",
    reportNumber = "HIP-2024-10/TH",
    doi = "10.1103/PhysRevD.111.075034",
    journal = "Phys. Rev. D",
    volume = "111",
    number = "7",
    pages = "075034",
    year = "2025"
}

@article{Kierkla:2025qyz,
    author = "Kierkla, Maciej and Schicho, Philipp and Swiezewska, Bogumila and Tenkanen, Tuomas V. I. and van de Vis, Jorinde",
    title = "{Finite-temperature bubble nucleation with shifting scale hierarchies}",
    eprint = "2503.13597",
    archivePrefix = "arXiv",
    primaryClass = "hep-ph",
    reportNumber = "CERN-TH-2025-046, HIP-2024-27/TH",
    doi = "10.1007/JHEP07(2025)153",
    journal = "JHEP",
    volume = "07",
    pages = "153",
    year = "2025"
}

@article{Camargo-Molina:2021zgz,
    author = {Camargo-Molina, Jos{\'e} Eliel and Enberg, Rikard and L{\"o}fgren, Johan},
    title = "{A new perspective on the electroweak phase transition in the Standard Model Effective Field Theory}",
    eprint = "2103.14022",
    archivePrefix = "arXiv",
    primaryClass = "hep-ph",
    doi = "10.1007/JHEP10(2021)127",
    journal = "JHEP",
    volume = "10",
    pages = "127",
    year = "2021"
}

@article{Camargo-Molina:2024sde,
    author = {Camargo-Molina, Eliel and Enberg, Rikard and L{\"o}fgren, Johan},
    title = "{A catalog of first-order electroweak phase transitions in the Standard Model Effective Field Theory}",
    eprint = "2410.23210",
    archivePrefix = "arXiv",
    primaryClass = "hep-ph",
    doi = "10.1007/JHEP08(2025)113",
    journal = "JHEP",
    volume = "08",
    pages = "113",
    year = "2025"
}

@article{Gould:2022ran,
    author = {Gould, Oliver and G{\"u}yer, Sinan and Rummukainen, Kari},
    title = "{First-order electroweak phase transitions: A nonperturbative update}",
    eprint = "2205.07238",
    archivePrefix = "arXiv",
    primaryClass = "hep-lat",
    reportNumber = "HIP-2022-10/TH",
    doi = "10.1103/PhysRevD.106.114507",
    journal = "Phys. Rev. D",
    volume = "106",
    number = "11",
    pages = "114507",
    year = "2022",
    note = "[Erratum: Phys.Rev.D 110, 119903 (2024)]"
}

@article{Brauner:2016fla,
    author = "Brauner, Tom{\'a}{\v{s}} and Tenkanen, Tuomas V. I. and Tranberg, Anders and Vuorinen, Aleksi and Weir, David J.",
    title = "{Dimensional reduction of the Standard Model coupled to a new singlet scalar field}",
    eprint = "1609.06230",
    archivePrefix = "arXiv",
    primaryClass = "hep-ph",
    reportNumber = "HIP-2016-27-TH",
    doi = "10.1007/JHEP03(2017)007",
    journal = "JHEP",
    volume = "03",
    pages = "007",
    year = "2017"
}

@article{Andersen:2017ika,
    author = "Andersen, Jens O. and Gorda, Tyler and Helset, Andreas and Niemi, Lauri and Tenkanen, Tuomas V. I. and Tranberg, Anders and Vuorinen, Aleksi and Weir, David J.",
    title = "{Nonperturbative Analysis of the Electroweak Phase Transition in the Two Higgs Doublet Model}",
    eprint = "1711.09849",
    archivePrefix = "arXiv",
    primaryClass = "hep-ph",
    reportNumber = "HIP-2017-26/TH, HIP-2017-26-TH",
    doi = "10.1103/PhysRevLett.121.191802",
    journal = "Phys. Rev. Lett.",
    volume = "121",
    number = "19",
    pages = "191802",
    year = "2018"
}

@article{Niemi:2018asa,
    author = "Niemi, Lauri and Patel, Hiren H. and Ramsey-Musolf, Michael J. and Tenkanen, Tuomas V. I. and Weir, David J.",
    title = "{Electroweak phase transition in the real triplet extension of the SM: Dimensional reduction}",
    eprint = "1802.10500",
    archivePrefix = "arXiv",
    primaryClass = "hep-ph",
    reportNumber = "HIP-2018-7-TH, ACFI-T18-04, HIP-2018-7/TH",
    doi = "10.1103/PhysRevD.100.035002",
    journal = "Phys. Rev. D",
    volume = "100",
    number = "3",
    pages = "035002",
    year = "2019"
}

@article{Gorda:2018hvi,
    author = "Gorda, Tyler and Helset, Andreas and Niemi, Lauri and Tenkanen, Tuomas V. I. and Weir, David J.",
    title = "{Three-dimensional effective theories for the two Higgs doublet model at high temperature}",
    eprint = "1802.05056",
    archivePrefix = "arXiv",
    primaryClass = "hep-ph",
    reportNumber = "HIP-2018-6/TH, HIP-2018-6-TH",
    doi = "10.1007/JHEP02(2019)081",
    journal = "JHEP",
    volume = "02",
    pages = "081",
    year = "2019"
}

@article{Kainulainen:2019kyp,
    author = "Kainulainen, Kimmo and Keus, Venus and Niemi, Lauri and Rummukainen, Kari and Tenkanen, Tuomas V. I. and Vaskonen, Ville",
    title = "{On the validity of perturbative studies of the electroweak phase transition in the Two Higgs Doublet model}",
    eprint = "1904.01329",
    archivePrefix = "arXiv",
    primaryClass = "hep-ph",
    doi = "10.1007/JHEP06(2019)075",
    journal = "JHEP",
    volume = "06",
    pages = "075",
    year = "2019"
}

@article{Gould:2019qek,
    author = "Gould, Oliver and Kozaczuk, Jonathan and Niemi, Lauri and Ramsey-Musolf, Michael J. and Tenkanen, Tuomas V. I. and Weir, David J.",
    title = "{Nonperturbative analysis of the gravitational waves from a first-order electroweak phase transition}",
    eprint = "1903.11604",
    archivePrefix = "arXiv",
    primaryClass = "hep-ph",
    reportNumber = "ACFI T19-04, HIP-2019-5/TH",
    doi = "10.1103/PhysRevD.100.115024",
    journal = "Phys. Rev. D",
    volume = "100",
    number = "11",
    pages = "115024",
    year = "2019"
}

@article{Gould:2021ccf,
    author = "Gould, Oliver and Hirvonen, Joonas",
    title = "{Effective field theory approach to thermal bubble nucleation}",
    eprint = "2108.04377",
    archivePrefix = "arXiv",
    primaryClass = "hep-ph",
    reportNumber = "HIP-2020-19/TH",
    doi = "10.1103/PhysRevD.104.096015",
    journal = "Phys. Rev. D",
    volume = "104",
    number = "9",
    pages = "096015",
    year = "2021"
}

@article{Gould:2021dzl,
    author = "Gould, Oliver",
    title = "{Real scalar phase transitions: a nonperturbative analysis}",
    eprint = "2101.05528",
    archivePrefix = "arXiv",
    primaryClass = "hep-ph",
    reportNumber = "HIP-2021-2/TH",
    doi = "10.1007/JHEP04(2021)057",
    journal = "JHEP",
    volume = "04",
    pages = "057",
    year = "2021"
}

@article{Schicho:2021gca,
    author = {Schicho, Philipp M. and Tenkanen, Tuomas V. I. and {\"O}sterman, Juuso},
    title = "{Robust approach to thermal resummation: Standard Model meets a singlet}",
    eprint = "2102.11145",
    archivePrefix = "arXiv",
    primaryClass = "hep-ph",
    doi = "10.1007/JHEP06(2021)130",
    journal = "JHEP",
    volume = "06",
    pages = "130",
    year = "2021"
}

@article{Lofgren:2021ogg,
    author = {L{\"o}fgren, Johan and Ramsey-Musolf, Michael J. and Schicho, Philipp and Tenkanen, Tuomas V. I.},
    title = "{Nucleation at Finite Temperature: A Gauge-Invariant Perturbative Framework}",
    eprint = "2112.05472",
    archivePrefix = "arXiv",
    primaryClass = "hep-ph",
    reportNumber = "ACFI-T21-15, HIP-2021-44/TH, NORDITA 2021-110",
    doi = "10.1103/PhysRevLett.130.251801",
    journal = "Phys. Rev. Lett.",
    volume = "130",
    number = "25",
    pages = "251801",
    year = "2023"
}

@article{Niemi:2021qvp,
    author = "Niemi, Lauri and Schicho, Philipp and Tenkanen, Tuomas V. I.",
    title = "{Singlet-assisted electroweak phase transition at two loops}",
    eprint = "2103.07467",
    archivePrefix = "arXiv",
    primaryClass = "hep-ph",
    reportNumber = "HIP-2021-8/TH, NORDITA 2021-011",
    doi = "10.1103/PhysRevD.103.115035",
    journal = "Phys. Rev. D",
    volume = "103",
    number = "11",
    pages = "115035",
    year = "2021",
    note = "[Erratum: Phys.Rev.D 109, 039902 (2024)]"
}

@article{Ekstedt:2022ceo,
    author = "Ekstedt, Andreas",
    title = "{Convergence of the nucleation rate for first-order phase transitions}",
    eprint = "2205.05145",
    archivePrefix = "arXiv",
    primaryClass = "hep-ph",
    doi = "10.1103/PhysRevD.106.095026",
    journal = "Phys. Rev. D",
    volume = "106",
    number = "9",
    pages = "095026",
    year = "2022"
}

@article{Ekstedt:2022zro,
    author = {Ekstedt, Andreas and Gould, Oliver and L{\"o}fgren, Johan},
    title = "{Radiative first-order phase transitions to next-to-next-to-leading order}",
    eprint = "2205.07241",
    archivePrefix = "arXiv",
    primaryClass = "hep-ph",
    doi = "10.1103/PhysRevD.106.036012",
    journal = "Phys. Rev. D",
    volume = "106",
    number = "3",
    pages = "036012",
    year = "2022",
    note = "[Erratum: Phys.Rev.D 110, 019901 (2024)]"
}

@article{Biondini:2022ggt,
    author = "Biondini, Simone and Schicho, Philipp and Tenkanen, Tuomas V. I.",
    title = "{Strong electroweak phase transition in t-channel simplified dark matter models}",
    eprint = "2207.12207",
    archivePrefix = "arXiv",
    primaryClass = "hep-ph",
    reportNumber = "HIP-2022-19/TH, NORDITA 2022-050",
    doi = "10.1088/1475-7516/2022/10/044",
    journal = "JCAP",
    volume = "10",
    pages = "044",
    year = "2022"
}

@article{Keus:2025ova,
    author = "Keus, Venus and Lewitt, Lucy and Thomson-Cooke, Jasmine",
    title = "{Electroweak phase transition enhanced by a CP-violating dark sector}",
    eprint = "2511.04636",
    archivePrefix = "arXiv",
    primaryClass = "hep-ph",
    reportNumber = "DIAS-STP-25-13",
    month = "11",
    year = "2025"
}

@article{Jahedi:2025yjz,
    author = "Jahedi, Sahabub and Saha, Indrajit and Sarkar, Abhik",
    title = "{Electroweak phase transition in SMEFT: Gravitational wave and collider complementarity}",
    eprint = "2512.04168",
    archivePrefix = "arXiv",
    primaryClass = "hep-ph",
    month = "12",
    year = "2025"
}

@article{Liu:2025ipj,
    author = "Liu, Jie and Qin, Renhui and Bian, Ligong",
    title = "{Gauge-independent treatment of electroweak phase transition}",
    eprint = "2512.05565",
    archivePrefix = "arXiv",
    primaryClass = "hep-ph",
    month = "12",
    year = "2025"
}

@article{Liu:2026ask,
    author = "Liu, Jie and Qin, Renhui and Bian, Ligong",
    title = "{Towards Accurate Gravitational Wave Predictions: Gauge-Invariant Nucleation in the Electroweak Phase Transition}",
    eprint = "2601.05793",
    archivePrefix = "arXiv",
    primaryClass = "hep-ph",
    month = "1",
    year = "2026"
}

@article{Chala:2025xlk,
    author = "Chala, Mikael and Fiore, Maria Cristina and Gil, Luis",
    title = "{Hot news on the phase-structure of the SMEFT}",
    eprint = "2507.16905",
    archivePrefix = "arXiv",
    primaryClass = "hep-ph",
    month = "7",
    year = "2025"
}

@article{Bernardo:2025vkz,
    author = "Bernardo, Fabio and Klose, Philipp and Schicho, Philipp and Tenkanen, Tuomas V. I.",
    title = "{Higher-dimensional operators at finite temperature affect gravitational-wave predictions}",
    eprint = "2503.18904",
    archivePrefix = "arXiv",
    primaryClass = "hep-ph",
    reportNumber = "HIP-2025-6/TH",
    doi = "10.1007/JHEP08(2025)109",
    journal = "JHEP",
    volume = "08",
    pages = "109",
    year = "2025"
}

@article{Chala:2025cya,
    author = "Chala, Mikael and Dashko, Andrii and Guedes, Guilherme",
    title = "{Running Couplings in High-Temperature Effective Field Theory}",
    eprint = "2510.26878",
    archivePrefix = "arXiv",
    primaryClass = "hep-ph",
    reportNumber = "CERN-TH-2025-220",
    month = "10",
    year = "2025"
}

@article{Chala:2025aiz,
    author = "Chala, Mikael and Guedes, Guilherme",
    title = "{The high-temperature limit of the SM(EFT)}",
    eprint = "2503.20016",
    archivePrefix = "arXiv",
    primaryClass = "hep-ph",
    doi = "10.1007/JHEP07(2025)085",
    journal = "JHEP",
    volume = "07",
    pages = "085",
    year = "2025"
}

@article{Bhatnagar:2025jhh,
    author = "Bhatnagar, Ansh and Croon, Djuna and Schicho, Philipp",
    title = "{Interpreting the 95 GeV resonance in the Two Higgs Doublet Model: Implications for the Electroweak Phase Transition}",
    eprint = "2506.20716",
    archivePrefix = "arXiv",
    primaryClass = "hep-ph",
    reportNumber = "IPPP/25/39",
    month = "6",
    year = "2025"
}

@article{Li:2025kyo,
    author = "Li, Xu-Xiang and Ramsey-Musolf, Michael J. and Tenkanen, Tuomas V. I. and Wu, Yanda",
    title = "{An Effective Sphaleron Awakens}",
    eprint = "2506.01585",
    archivePrefix = "arXiv",
    primaryClass = "hep-ph",
    reportNumber = "HIP-2025-8/TH",
    month = "6",
    year = "2025"
}

@article{Annala:2025aci,
    author = "Annala, Jaakko and Rummukainen, Kari and Tenkanen, Tuomas V. I.",
    title = "{Nonperturbative determination of the sphaleron rate for first-order phase transitions}",
    eprint = "2506.04939",
    archivePrefix = "arXiv",
    primaryClass = "hep-ph",
    reportNumber = "HIP-2025-19/TH",
    doi = "10.1103/q1jq-gq9m",
    journal = "Phys. Rev. D",
    volume = "113",
    number = "1",
    pages = "016014",
    year = "2026"
}

@article{Gould:2024jjt,
    author = "Gould, Oliver and Saffin, Paul M.",
    title = "{Perturbative gravitational wave predictions for the real-scalar extended Standard Model}",
    eprint = "2411.08951",
    archivePrefix = "arXiv",
    primaryClass = "hep-ph",
    doi = "10.1007/JHEP03(2025)105",
    journal = "JHEP",
    volume = "03",
    pages = "105",
    year = "2025"
}

@article{Chakrabortty:2024wto,
    author = "Chakrabortty, Joydeep and Mohanty, Subhendra",
    title = "{One Loop Thermal Effective Action}",
    eprint = "2411.14146",
    archivePrefix = "arXiv",
    primaryClass = "hep-th",
    doi = "10.1016/j.nuclphysb.2025.117165",
    journal = "Nucl. Phys. B",
    volume = "1020",
    pages = "117165",
    year = "2025"
}

@article{Laine:2019uua,
    author = {Laine, M. and Schicho, P. and Schr{\"o}der, Y.},
    title = "{A QCD Debye mass in a broad temperature range}",
    eprint = "1911.09123",
    archivePrefix = "arXiv",
    primaryClass = "hep-ph",
    doi = "10.1103/PhysRevD.101.023532",
    journal = "Phys. Rev. D",
    volume = "101",
    number = "2",
    pages = "023532",
    year = "2020"
}

@article{Laine:2018lgj,
    author = {Laine, M. and Schicho, P. and Schr{\"o}der, Y.},
    title = "{Soft thermal contributions to 3-loop gauge coupling}",
    eprint = "1803.08689",
    archivePrefix = "arXiv",
    primaryClass = "hep-ph",
    doi = "10.1007/JHEP05(2018)037",
    journal = "JHEP",
    volume = "05",
    pages = "037",
    year = "2018"
}

@article{Gorda:2025cwu,
    author = {Gorda, Tyler and Navarrete, Pablo and Paatelainen, Risto and Sandbote, Leon and Sepp{\"a}nen, Kaapo},
    title = "{A new approach to determine the thermodynamics of deconfined matter to high accuracy}",
    eprint = "2511.09627",
    archivePrefix = "arXiv",
    primaryClass = "hep-ph",
    reportNumber = "HIP-2025-32/TH",
    month = "11",
    year = "2025"
}

@article{Ghiglieri:2021bom,
    author = "Ghiglieri, Jacopo and Moore, Guy D. and Schicho, Philipp and Schlusser, Niels",
    title = "{The force-force-correlator in hot QCD perturbatively and from the lattice}",
    eprint = "2112.01407",
    archivePrefix = "arXiv",
    primaryClass = "hep-ph",
    reportNumber = "HIP-2021-43/TH",
    doi = "10.1007/JHEP02(2022)058",
    journal = "JHEP",
    volume = "02",
    pages = "058",
    year = "2022"
}

@article{Biekotter:2025npc,
    author = {Biek{\"o}tter, Thomas and Dashko, Andrii and L{\"o}schner, Maximilian and Weiglein, Georg},
    title = "{Perturbative aspects of the electroweak phase transition with a complex singlet and implications for gravitational wave predictions}",
    eprint = "2511.14831",
    archivePrefix = "arXiv",
    primaryClass = "hep-ph",
    reportNumber = "DESY-25-131, IFT-UAM/CSIC-25-104",
    month = "11",
    year = "2025"
}

@article{Karkkainen:2025nkz,
    author = {K{\"a}rkk{\"a}inen, Aapeli and Navarrete, Pablo and Nurmela, Mika and Paatelainen, Risto and Sepp{\"a}nen, Kaapo and Vuorinen, Aleksi},
    title = "{Quark Matter at Four Loops: Hardships and How to Overcome Them}",
    eprint = "2501.17921",
    archivePrefix = "arXiv",
    primaryClass = "hep-ph",
    reportNumber = "HIP-2025-01/TH",
    doi = "10.1103/627n-5g6l",
    journal = "Phys. Rev. Lett.",
    volume = "135",
    number = "2",
    pages = "021901",
    year = "2025"
}

@article{Fuentes-Martin:2024agf,
    author = "Fuentes-Mart{\'\i}n, Javier and Moreno-S{\'a}nchez, Adri{\'a}n and Palavri{\'c}, Ajdin and Thomsen, Anders Eller",
    title = "{A guide to functional methods beyond one-loop order}",
    eprint = "2412.12270",
    archivePrefix = "arXiv",
    primaryClass = "hep-ph",
    doi = "10.1007/JHEP08(2025)099",
    journal = "JHEP",
    volume = "08",
    pages = "099",
    year = "2025"
}

@article{DeAngelis:2025agn,
    author = "De Angelis, Stefano and Kosower, David A. and Ma, Rourou and Wu, Zihao and Zhang, Yang",
    title = "{Singularity-free Feynman integral bases}",
    eprint = "2508.04394",
    archivePrefix = "arXiv",
    primaryClass = "hep-th",
    reportNumber = "USTC-ICTS/PCFT-25-30, MPP-2025-151",
    doi = "10.1103/zd34-cc4y",
    journal = "Phys. Rev. D",
    volume = "113",
    number = "5",
    pages = "056013",
    year = "2026"
}

\end{document}